\documentclass[titlepage,12pt]{article}
\usepackage[margin=1in]{geometry}
\usepackage{latexsym}
\usepackage{amsmath}
\usepackage{amsfonts}
\usepackage{amsthm}
\usepackage{amssymb}
\usepackage{mathstyle}
\usepackage{hyperref,setspace}
\usepackage{natbib}

\usepackage{graphicx}
\newcommand{\bl}{\begin{flushleft}}
\newcommand{\el}{\end{flushleft}}

\newcommand{\D}{{d}}
\newcommand{\re}{{\mathbb{R}}}
\newcommand{\pn}{{\mathbb{P}_n}}

\newcommand{\bX}{{\boldsymbol X}}
\newcommand{\bx}{{\boldsymbol x}}

\newcommand{\cond}{|}

\newcommand{\argmin}{\mbox{argmin}}
\newcommand{\sign}{ \mathrm{sgn}}
 
\newcommand{\R}{\mathcal{R}}

\newtheorem{remark}{Remark}
\hypersetup{linkcolor=blue,citecolor=black,filecolor=cyan,urlcolor=black}

\newtheorem{lem}{Lemma}[section]
\newtheorem{prop}{Proposition}[section]
\newtheorem{thm}{Theorem}[section]
\newtheorem{cor}{Corollary}[section]
\newtheorem{assumption}{Assumption}

\begin{document}

\title{Efficient augmentation and relaxation learning for
   individualized treatment rules using observational data}

\author{Ying-Qi
   Zhao\thanks{ Public Health Sciences Division, Fred Hutchinson Cancer Research Center, Seattle, WA, 98109, Email: yqzhao@fhcrc.org.} and Eric
   B. Laber\thanks{  Department of Statistics,
     North Carolina State University, Raleigh, NC 27695 
} and Yang Ning\thanks{ Department of Statistical Science,
	   Cornell University, Ithaca, NY, 14853 } \\and Sumona Saha\thanks{  Department of Medicine,
     University of Wisconsin, Madison, WI,  53705   
} and Bruce E. Sands\thanks{  Division of Gastroenterology, Icahn School of Medicine at  Mount Sinai, New York, NY, 10029  }}
\date{}

%  This label and the label ``lastpage'' are used by the \pagerange
%  command above to give the page range for the article

\maketitle 

\begin{abstract}
 Individualized treatment rules aim to identify if, when, which, and to whom treatment should be applied. A globally aging population,  rising healthcare costs, and increased access to patient-level data  have created an urgent need for high-quality estimators of  individualized treatment rules that can be applied to observational  data.  A recent and promising line of research for estimating  individualized treatment rules recasts the problem of  estimating an optimal treatment rule as a weighted classification  problem.  We consider a class of estimators for optimal  treatment rules that are analogous to convex large-margin
  classifiers.  The proposed class applies to observational data and  is doubly-robust in the sense that correct specification of either a  propensity or outcome model leads to consistent estimation of the  optimal individualized treatment rule.  Using techniques from semiparametric efficiency theory, we derive rates of convergence for the proposed estimators
  and use these rates to characterize the bias-variance trade-off for  estimating individualized treatment rules with  classification-based methods.  Simulation experiments informed by  these results demonstrate that it is possible to construct new  estimators within the proposed framework that significantly  outperform existing ones.  We illustrate the proposed methods  using data from a labor training program and a study of inflammatory  bowel syndrome.
\end{abstract}

\maketitle

\section{Introduction}\label{scnintro}
There is a growing consensus that the best possible care results from
treatment decisions that are carefully tailored to individual patient
characteristics \citep[][]{sox2009comparative}.  Individualized
treatment rules (ITRs) 
%also known as treatment selection rules \citep[][]{yingBIOST},
formalize tailored treatment decisions as a function from patient
information to a recommended treatment.  We define an optimal ITR
as maximizing the mean of a pre-specified clinical outcome if applied to
recommend treatments in a population of interest \citep[see][for
alternative
definitions of optimality]{linn2016interactive}.  With expanding
access to patient-level data through electronic health records,
adverse event reporting,  insurance claims, and billing
records, there is increasing interest in estimating optimal
ITRs from observational data.  An important use of an estimated
optimal ITR is hypothesis-generation whereby the
estimated optimal rule is used to discover covariate-treatment
interactions or identify subgroups of patients with large treatment
effects.  In such applications, it is useful to directly control the
class of ITRs within which the optimal ITR will be estimated.
The form of this class can be chosen to ensure interpretability,
enforce logistical or cost constraints, or make the tests of certain
clinical hypotheses overt.

One approach to estimating an optimal ITR is to model some or all of
the conditional distribution of the outcome given treatments and
covariates and then to use this estimated distribution to infer the
optimal ITR.  These approaches are sometimes called indirect methods
as they indirectly specify the form of the optimal ITR through
postulated models for components of the conditional outcome
distribution.  Indirect methods have dominated the literature on
estimating optimal ITRs; examples of indirect estimation methods
include variations of $g$-estimation in structural nested models
\citep[][]{robins1989analysis, Robins:CausalNotes1997,
  Murphy03optimaldynamic, Robins04optimalstructural}; $Q$- and
$A$-learning \citep[][]{Zhao:RL2009, qian:itr11, moodie2012q,
  bibhasBook,  Schulte:QandA2012}, and regret regression
\citep[][]{henderson2009regret}.  However, a major drawback with these
approaches is that the postulated outcome models dictates the class of
possible ITRs.  A consequence is that to obtain a simple ITR
 requires specification of simple outcome models, which may not
be correctly specified.  Moreover, if these outcome models are
misspecified, the foregoing methods may not be consistent for the
optimal ITR within the class implied by the outcome models.  For
example, to ensure a linear ITR using $Q$-learning, it is common to use
a linear conditional mean model.  It can be shown that if the
linear mean model is misspecified then the estimated optimal ITR using
$Q$-learning need not converge to the optimal linear ITR
\citep[][]{qian:itr11}.  Alternatively, flexible outcome models that
mitigate the risk of misspecification \citep[e.g.,][]{Zhao:RL2009,
  qian:itr11, moodie2013q} can induce a class of ITRs that is
difficult or impossible to interpret (see Section 2 for details).

An alternative to indirect estimation is to decouple models for the
conditional outcome distribution from the class of ITRs.  One way
to do this is to form a flexible estimator of the mean outcome as a
function of the ITR that is consistent under a large
class of potential generative models and then to use the maximizer of
this function over a pre-specified class of ITRs as the estimator
of the optimal ITR.  These approaches are called direct
\citep[][]{laber2014dynamic}, policy-search
\citep[][]{Sutton:Reinforcementlearning98, szepesvari2010algorithms}, policy learning \citep{athey2017efficient} 
or value-search \citep[][]{marieChapter} estimators.  An advantage of
direct estimators is that they permit flexible, e.g., semi- or
non-parametric, models for modeled portions of the outcome
distribution yet still control the form of the estimated optimal ITR.  Direct
estimators include outcome weighted learning (Zhao et al., 2012, 2015a, 2015b), robust value-search estimators (Zhang et al., 2012a, 2012b, 2013); marginal
structural mean models \citep[][]{robinsetal2008, Orellana10};
and Q-learning with policy-search
\citep[][]{taylor2015reader, zhang2015using, zhang2017estimation}.  

While the foregoing methods represent significant progress in direct
estimation, computational and theoretical gaps remain.  Outcome
weighted learning uses a convex relaxation of an inverse-probability
weighted estimator (IPWE) of the mean outcome.  This convex relaxation
makes their method computationally efficient and scalable to large
problems; in addition, convexity simplifies derivations of convergence
rates and generalization error bounds.  However, the IPWE is known to
be unstable under certain generative models (Zhang et al., 2012a,
2012b), and theoretical guarantees for outcome weighted learning were
developed only for data from a randomized clinical trial.  Robust
value-search estimators directly maximize an augmented IPWE (AIPWE).
The AIPWE is semi-parametric efficient and is significantly more
stable than the IPWE.  However, the AIPWE is a discontinuous function
of the observed data, which makes direct maximization computationally
burdensome even in moderate sized problems and complicates theoretical
study of these estimators. We establish the theory for both AIPW and
its convex relaxation, which fills the gap in the current literature
on direct search methods. Marginal structural mean models are best
suited for problems where the ITR depends only on a very small number
of covariates. \citet{liu2016robust} proposed a robust method for
estimating optimal treatment rules in a multi-stage setup. At each
stage in a multi-stage setup, they proposed a robust weight to replace
the original weight in OWL based on the idea of augmentation. However,
they still require consistent estimation of the propensity score at
the present stage. In particular, their proposal for the single stage
problem still relies on an IPWE, and does not possess the double
robustness property.

We propose a class of estimators representable as the maximizer of a
convex relaxation of the AIPWE; we term this class of estimators
Efficient Augmentation and Relaxation Learning (EARL).  EARL is
computationally efficient, theoretically tractable, and applies to
both observational and experimental data.  Furthermore, EARL contains
outcome weighted learning (OWL) (Zhao et al., 2012)
as a special case.  
However, EARL is considerably more general than OWL,
and this generality leads to new insights about classification-based
estimation of ITRs, new algorithms, and new theoretical results.
%The contributions of this work are as follows.    
Unlike OWL, EARL makes use of both a propensity score and an outcome
regression model.  Estimators within the EARL framework are
doubly-robust in the sense that they consistently estimate the optimal
ITR if either the propensity score model or outcome regression model
is correctly specified.  Within the EARL framework, we are able to
characterize convergence rates across a range of convex relaxations,
propensity score models, and outcome regression models. In particular,
making use of sample splitting, we are able to remove the dependence
in estimating the nuisance functions and in constructing the estimated
ITR. We show that under all convex relaxations considered, a fast
convergence rate of the estimated optimal ITR can be achieved, and
that the estimation of the propensity score and outcome regression
models need not affect the upper bound of this rate.  The proposed
method has been implemented in R and is freely available through the
`DynTxRegime' package hosted on the comprehensive R network
(cran.org).
%We show that under all convex relaxations considered, a sub-parametric convergence rate of either the outcome or propensity model is associated with a sub-parametric convergence rate of the estimated optimal ITR. 
  
In Section 2, we introduce the EARL class of estimators. 
In Section 3, we investigate the theoretical properties of
estimators within this class.  
In Section 4, we use simulation experiments to investigate
the finite sample performance of EARL estimators.  
In Section 5, we present illustrative case studies using
data from a labor training program and an inflammatory bowel disease
study.  In Section 6, we make concluding remarks and discuss potential
extensions.

\section{Methods}\label{scnmethods}
\subsection{Background and preliminaries}
%\subsection{Value function and individualized treatment rules}
The observed data, $\lbrace (\bX_i, A_i, Y_i)\rbrace_{i=1}^{n}$,
comprise $n$ independent, identically distributed copies of
$(\bX, A, Y)$, where: $\bX \in \mathbb{R}^{p}$ denotes baseline
subject measurements; $A\in\lbrace-1,1\rbrace$ denotes the assigned
treatment; and $Y\in\mathbb{R}$ denotes the outcome, coded so that
higher values are better.  In this context, an ITR, $d$, is a map
from $\mathbb{R}^p$ into $\lbrace -1, 1\rbrace$ so that a patient
presenting with $\bX =\bx$ is recommended treatment $d(\bx)$.  Let
$\mathcal{D}$ denote a class of ITRs of interest.  To define the
optimal ITR, denoted $d^{*}$, we use the framework of potential
outcomes \citep[][]{Rubin:Causal1974, Neyman:PotentialOutcome1990}.
Let $Y(a)$ denote the potential outcome under treatment
$a\in\lbrace -1,1\rbrace$ and define
$Y(d) = \sum_{a\in\lbrace-1,1\rbrace}Y(a)I\lbrace a=d(\bX)\rbrace$ to
be the potential outcome under $d$.  The marginal mean outcome
$V(d) \triangleq E\{Y(d)\}$ is called the value of the ITR $d$.  The
optimal ITR satisfies $d^*\in\mathcal{D}$ and $V(d^*) \ge V(d)$ for
all $d\in\mathcal{D}$.  Note that this definition of optimality depends on
the class $\mathcal{D}$.  To express the value in terms of the data
generating model, we assume: (i) strong ignorability,
$\{Y(-1), Y(1)\} \amalg A \big| \bX$
\citep[][]{Rubin:Causal1974, Robins:Causal1986,
  Neyman:PotentialOutcome1990}; (ii) consistency, $Y = Y(A)$; and
(iii) positivity, there exists $\tau > 0$ so that $\tau < P(A=a|\bX)$
for each $a\in\left\lbrace -1,1\right\rbrace$ with probability one.
These assumptions are common and well-studied \citep[see][for a recent
review of potential outcomes for treatment
rules]{Schulte:QandA2012}.  Assumption (i) is true in a randomized
study but unverifiable in an observational study \citep[][]{bang2005doubly}.

Define $Q(\bx ,a) \triangleq E(Y|\bX= \bx, A=a)$, then under the
foregoing assumptions, it can be shown that
\begin{equation}\label{refQFn}
V(d) = E\left[ Q\left\lbrace \bX,
        d(\bX)\right\rbrace\right],
\end{equation}
from which it follows 
that $d^*(\bx) = \arg\max_{a\in\lbrace -1,1\rbrace}Q(\bx, a)$.
Q-learning is a common regression-based indirect approach 
for estimating $d^*$ wherein an estimator  $\widehat{Q}(\bx, a)$ of
$Q(\bx, a)$ 
is constructed and subsequently the estimated optimal
rule is $\widehat{d}(\bx) =\arg\max_{a}\widehat{Q}(\bx, a)$.  
%Indirect estimators typically model either the $Q$-function,
%$Q(\bx, a)$, or the regret function,
%$\{\max_{a'\in\lbrace -1,1\rbrace}Q(\bx, a')\}-Q(\bx, a)$.  
%To illustrate some of the points about indirect estimation  made in Section \ref{scnintro}, we use $Q$-learning, which estimates the $Q$-function by estimating the regression of $Y$ on $\bX$ and $A$. Let $\mathcal{Q} = \left\lbrace \widetilde Q(\bx,  a;\theta)\,:\,\theta\in\Theta \right\rbrace$ denote a postulated class of models for the $Q$-function. $Q$-learning first constructs an estimator of the $Q$-function  within $\mathcal{Q}$, say $\widetilde Q(\bx, a;\widehat{\theta})$, and subsequently constructs the estimator  $\widehat{d}(\bx) = \arg\max_{a\in\lbrace -1,1\rbrace} \widetilde Q(\bx, a;\widehat{\theta})$ of the optimal ITR. Thus, the ITR estimated by $Q$-learning must be of the form $\arg\max_{a\in\lbrace -1,1\rbrace} \widetilde Q(\bx, a;\theta)$ for some $\theta\in\Theta$;  the postulated class of $Q$-functions dictates the class of rules.  However, it can be shown that even if $d^*(\bx) = \arg\max_{a}\widetilde Q(\bx, a;\theta^*)$ for some $\theta^*\in\Theta$ the estimated rule $\widehat{d}$ need not be consistent for $d^*$ if $Q(\bx, a) \notin \mathcal{Q}$ \citep{qian:itr11}.  
Let $\mathcal{Q}$ denote the postulated class of models for $Q(\bx,
a)$, then the set of possible decision rules obtained using
$Q$-learning is $\mathcal{D} = \left\lbrace d\,:\,
  d(\bx)=\arg\max_{a}Q(\bx, a), \, Q\in\mathcal{Q}\right\rbrace$.  
Thus, there is an inherent trade-off between choosing 
$\mathcal{Q}$ to be sufficiently rich to reduce the risk of model
misspecification and the resultant complexity of the resultant 
class of ITRs.

Direct estimators specify a class of candidate ITRs independently 
from postulated models for some or all of the generative model. 
Let $\mathcal{D}$ denote a class of ITRS; direct search estimators 
%Define the value function
%$V:\mathcal{D}\rightarrow\mathbb{R}$ to be the map $d\mapsto V(d)$.
first construct an estimator of the value function, say
$\widehat{V}(\cdot)$, and then choose
$\widehat{d} = \arg\max_{d\in\mathcal{D}}\widehat{V}(d)$ as the
estimator of $d^*$.  Thus, a
complex model space for $V(\cdot)$ need not imply a complex class of
rules $\mathcal{D}$.  However, the class of models for $V(\cdot)$
must be sufficiently rich to avoid implicit, unintended restrictions
on $\widehat{d}$.  %For example, suppose $\mathcal{D}_{\mathrm{Quad}}$ is the class of quadratic treatment rules, $\mathcal{D}_{\mathrm{Quad}} =\left\lbrace d \in \lbrace  -1,1\rbrace^{\mathbb{R}^p}\,:\, d(\bx) = \mathrm{sgn}\left(\beta_0 +    \bx^{T}\beta + \bx^{T}B\bx \right),\,\beta_0\in\mathbb{R},\,  \beta\in\mathbb{R}^{p}, B\in\mathbb{R}^{p\times p} \right\rbrace$ but the class of working models for $V(\cdot)$ is $\mathcal{V}_{\mathrm{Lin}} = \big\lbrace V\in\mathbb{R}^{\mathcal{D}}\,:\, V(d) = \beta_0 + E\left\lbrace d(\bX)\bX^{T}\beta\right\rbrace,\, \beta_0\in\mathbb{R}, \,\beta \in \mathbb{R}^{p}\big\rbrace$. Then, for any $V\in\mathcal{V}_{\mathrm{Lin}}$, the rule $\arg\max_{d\in\mathcal{D}_{\mathrm{Quad}}\, and\, V(d)\in \mathcal{V}_{Lin}}V(d)$ must be linear as $\arg\max_{d\in\mathcal{D}_{\mathrm{Quad}}}\left[ \beta_0 +  E\left\lbrace d(\bX)\bX^{T}\beta\right\rbrace \right]$ is $d(\bx) = \mathrm{sgn}(\bx^{T}\beta)$.  Thus, $\bigcup_{V\in\mathcal{V}_{Lin}}\arg\max_{d\in\mathcal{D}_{\mathrm{Quad}}}V(d) \subsetneq \mathcal{D}_{\mathrm{Quad}}$.
%The model space for $V(\cdot)$, if not sufficiently expressive,
%can restrict the form of the estimated optimal rule. 
 To avoid such
restrictions and to avoid model-misspecification, it is common to use a
flexible class of semi- or non-parametric models for $V(\cdot)$.

\subsection{Augmentation for the value function}\label{scn:aug}
%Direct-search estimators maximize an estimator of (\ref{valuefcn}) over a predefined class of regimes $\mathcal{D}$. 
Define the propensity score $\pi(a;\bx) \triangleq P(A=a|\bX=\bx)$,
then
\begin{equation}
   V(d) = E\left[\frac{Y}{\pi(A; \bX)}I\{A=d(\bX)\}\right],
   \label{valuefcn}
\end{equation}
where $I\{\cdot\}$ denotes the indicator function
\cite[e.g.,][]{qian:itr11}.  Unlike (\ref{refQFn}), the preceding
expression does not require an estimator of the $Q$-function.  Given
an estimator of the propensity score, $\widehat{\pi}(a;\bx)$, a
plug-in estimator for $V(d)$ based on (\ref{valuefcn}) is the inverse
probability weighted estimator (IPWE)
$\widehat{V}^{\mathrm{IPWE}}(d) \triangleq \pn [Y I\left\lbrace
  A=d(\bX)\right\rbrace/\widehat{\pi}(A;X)]$,
where $\pn$ is the empirical distribution. The IPWE has potentially
high variance as it only uses outcomes from subjects
whose treatment assignments coincide with those recommended by $d$.

One
approach to reduce variability is to augment the IPWE with a 
term involving both the propensity score and the $Q$-function that
is estimated using data from all of the observed subjects
\citep{robins1994estimation,cao2009improving}. 
Let $\widehat{Q}(\bx, a)$ denote an estimator of $Q(\bx,a)$. 
The augmented inverse probability weighted estimator  is
\begin{equation}
 \widehat V^{\mathrm{AIPWE}}(d)  \triangleq \pn\left[\frac{YI\left\lbrace
  A=d(\bX)
\right\rbrace}{\widehat{\pi}\left\lbrace d(X) ;X \right\rbrace} - 
\frac{I\left\lbrace A=d(\bX)\right\rbrace -  \widehat{\pi}\{d(\bX);\bX\}}
{\widehat{\pi}\{d(\bX);\bX\}}
\widehat{Q}\left\lbrace \bX, d(\bX)\right\rbrace
\right]. 
\label{aipwe}
\end{equation} 
It can be seen that $\widehat V^{\mathrm{AIPWE}}(d)$ is equal to
$\widehat V^{\mathrm{IPWE}}(d)$ plus an estimator of zero built using
outcomes from all subjects regardless of  whether or not their treatment
assignment is consistent with $d$.  If $\widehat{Q}(\bx, a) \equiv 0$ 
then $\widehat{V}^{\mathrm{AIPWE}}(d)=\widehat{V}^{\mathrm{IPWE}}(d)$ for 
all $d$.  

Hereafter, we use $\widehat{Q}(\bx, a)$ and $\widehat{\pi}(a;\bx)$ 
to denote generic estimators of the $Q$-function and propensity score.
The following assumption is used to establish double 
robustness of $\widehat{V}^{\mathrm{AIPWE}}(d)$.  
\begin{assumption}
\label{assp_limit}
$\widehat Q(\bx,a)$ and $\widehat{\pi}(a;\bx)$ converge in probability
uniformly to deterministic limits $Q^{m}(\bx, a)$  and
$\pi^{m}(a;\bx)$.
\end{assumption}\noindent
% The second term in (\ref{aipwe}) is to reduce variability produce
% robustness against model misspecification \citep[see
% also][]{zhang2012robust}.
This assumption does not require that the estimators
$\widehat{Q}(\bx, a)$, $\widehat{\pi}(a;\bx)$ are consistent for the truth, only that
they converge to fixed functions.
The following result is proved in 
Web Appendix A.
\begin{lem}\label{lemaug}
  Let $d\in\mathcal{D}$ be fixed. If {\em either} $\pi^m(a;\bx)=\pi(a;\bx)$  
  or $Q^m(\bx, a)=Q(\bx, a)$ for all $(\bx, a)$ outside of 
a set of measure zero, then $\widehat V^{AIPWE}(d)\rightarrow_{p} 
V^{AIPWE,m}(d)=V(d)$, where
\begin{equation*}
V^{AIPWE,m}(d) \triangleq E\left[
\frac{YI\left\lbrace
  A=d(\bX)
\right\rbrace}{{\pi}^{m}(A;\bX)} - 
\frac{I\left\lbrace A=d(\bX)\right\rbrace -  {\pi}^{m}\{d(\bX);\bX\}}
{{\pi}^{m}\{d(\bX);\bX\}}{Q}^{m}\{\bX, d(\bX)\}
\right].
\end{equation*}
\end{lem}\noindent
The preceding result shows that $\widehat{V}^{AIPWE}(d)$ is
doubly-robust in the sense that if either the propensity model or the
modeled $Q$-function is consistent, but not necessarily both,  then
$\widehat{V}^{AIPWE}(d)$ is consistent for $V(d)$. Thus, the
maximizer of $\widehat{V}^{AIPWE}(d)$ over $d\in\mathcal{D}$ is termed
a doubly-robust estimator of the optimal treatment rule
(Zhang et al., 2012a, 2012b, 2013).  However,
because $\widehat{V}^{AIPWE}(d)$ is not continuous, computing this
doubly-robust estimator can be computationally infeasible even in
moderate problems \citep[][]{baqun2012}.
%, e.g., $p \ge 10$ and $\mathcal{D}$ the set of
%linear rules.  
Instead, we form an estimator by maximizing a concave
relaxation of $\widehat{V}^{AIPWE}(d)$.  Maximizing this concave
relaxation is computationally efficient even in very high-dimensional
problems.  We show that the maximizer of this relaxed criteria remains
doubly-robust. Furthermore, we show that 
the rates of convergence of the proposed estimators
depend on the chosen concave relaxation, the chosen
propensity model, and the chosen model for the $Q$-function.  The
relationships among these choices provides
new knowledge about direct search estimators based on concave
surrogates (Zhang et al., 2012; Zhao et al., 2012, 2015a, 2015b).

\subsection{Efficient augmentation and relaxation learning (EARL)}
 
Let $\mathcal{M}$ be the class of measurable functions from
$\mathbb{R}^p$ into $\mathbb{R}$. Any decision rule $d(\bx)$ can be
written as $d(\bx) = \mathrm{sgn}\{f(\bx)\}$ for some function
$f \in \mathcal{M}$, where we define $\mathrm{sgn}(0) = 1$.
% If the function
% $\bx\mapsto \mathrm{sgn}\left\lbrace 
% Q(\bx, 1)-Q(\bx, -1)\right\rbrace$ belongs to $\mathcal{D}$ then 
% $d^*(\bx) = \mathrm{sgn}\left\lbrace 
% Q(\bx, 1)-Q(\bx, -1)\right\rbrace$.
 %The class of regimes implied by $\mathcal{F}$ is $\mathcal{D} = \left\lbrace d(\bx) = \mathrm{sgn}\left\lbrace f(\bx)\right\rbrace \,:\, f\in\mathcal{F}\right\rbrace$.
For $d(\bx) = \mathrm{sgn}\{f(\bx)\}$, $I\lbrace
a=d(\bx)\rbrace = I\lbrace af(\bx) \ge 0\rbrace$. Define $V(f)$,
${V}^{\mathrm{IPWE},m}(f)$, and ${V}^{\mathrm{AIPWE},m}(f)$ by substituting $I\lbrace A
f(\bX) \ge 0 \rbrace$ for $I\lbrace A = d(\bX)\rbrace$ in their
respective definitions. 
Define 
%\begin{eqnarray*}
% $W_a(Y,\pi,h) = I(A=a)\frac{Y-h(\bX,a)}{\pi(a;\bX)}+h(\bX,a)$,\,a=\pm 1 . 
%\end{eqnarray*}
$$W_a^m=W_a(Y, \bX, A, \pi^m, Q^m)=\frac{YI(A=a)}{\pi^m(a;\bX)} - 
\frac{I(A=a)-\pi^m(a; \bX)}{\pi^m(a;\bX)}Q^m(\bX,a), a\in\lbrace -1,1\rbrace.$$ 
The following result shows that 
maximizing $\widehat{V}^{\mathrm{AIPWE}}(f)$ is equivalent to minimizing 
a sum of weighted misclassification rates; a proof is given in 
Web Appendix B.
\begin{lem}\label{lemma_trans}
 Assume that $P\{f(\bX) = 0\} = 0$. Define $\widehat{f}_{n} =
  \arg\sup_{f\in\mathcal{M}}\widehat{V}^{\mathrm{AIPWE}}(f)$, then
  \begin{equation*}
    \widehat{f}_{n} = \arg\inf_{f\in\mathcal{M}}\pn \left[
      |\widehat W_{1}|
    I\left\lbrace \mathrm{sgn}
      (\widehat W_{1}) f(\bX) < 0\right\rbrace
     + |\widehat W_{-1}|
    I\left\lbrace -\mathrm{sgn}(\widehat W_{-1}) f(\bX) < 0
      \right\rbrace
    \right],
  \end{equation*}
  where $ \widehat W_a = W_a(Y, \bX, A, \widehat \pi, \widehat{Q}), a=\in 
\lbrace -1, 1\rbrace $.
%  \begin{equation*}
%    \widehat W_a^m =I(A=a) \frac{Y-\widehat{Q}^{m}(\bX, j)}{\widehat{\pi}^{m}(a;\bX)}
%    + \widehat{Q}^{m}(\bX, j), j = -1, 1.
%  \end{equation*}
\end{lem}

\noindent
Lemma \ref{lemma_trans} shows that the estimator, $\widehat{f}_{n}$,
which maximizes $\widehat{V}^{\mathrm{AIPWE}}(f)$ over
$f\in\mathcal{M}$, can be viewed as minimizing a sum of weighted
$0$-$1$ losses.  In this view, the class labels are
$\mathrm{sgn}(\widehat W_a)\cdot a$ and the misclassification weights
are $|\widehat W_a|$, $a=\in \lbrace -1,1\rbrace $
\citep[see][]{zhang2012robust, zhang2013robust}.  Directly minimizing
the combined weighted $0$-$1$ loss is a difficult non-convex
optimization problem \citep[][]{laber2011adaptive}.  One strategy to
reduce computational complexity is to replace the indicator function
with a convex surrogate and to minimize the resulting relaxed
objective function \citep[][]{freund1999large, bartlett:convexity06,
  hastie:esl09}.  This strategy has proved successful empirically and
theoretically in classification and estimation of optimal treatment
rules \citep[][]{Zhao:OWL12}. However, unlike previous applications of
convex relaxations to the estimation of optimal treatment rules, we
establish rates of convergence as a function of the: (i) choice of
convex surrogate; (ii) convergence rate of the postulated propensity
score estimator; and (iii) convergence rate the postulated
$Q$-function estimator. We characterize the relationship among these
three components in Section \ref{scntheory}.

The function $f$ is conceptualized as being a smooth function of $\bx$
that is more easily constrained to possess certain desired structure,
e.g., sparsity, linearity, etc. Thus, we will focus on estimation of
$f$ within a class of functions $\mathcal{F}$ called the approximation
space; we assume that $\mathcal{F}$ is a Hilbert space with norm
$\|\cdot\|$. %that contains $f^*$.
Let $\phi:\mathbb{R}\rightarrow \mathbb{R}$ denote a convex function and
define EARL
estimators as those taking the form
\begin{equation}
  \tilde{f}_{n}^{\lambda_{n}} = \arg\inf_{f\in\mathcal{F}}\pn \left[|\widehat W_1|\phi\left\{
    \mathrm{sgn}(\widehat W_1)f(\bX)
    \right\} + |\widehat W_{-1}|\phi\left\{
    -\mathrm{sgn}(\widehat W_{-1})f(\bX)
    \right\} \right]+ \lambda_{n} \|f\|^{2},
\label{hatFL}
\end{equation}
where $\lambda_{n}\|f\|^2$ is included to reduce
overfitting and $\lambda_{n} \ge 0$ is a (possibly data-dependent)
tuning parameter.  Throughout, we assume that $\phi(t)$ is one of the
following:
hinge loss, $\phi(t)=\max(1-t,0)$; exponential loss, $\phi(t)=e^{-t}$;
logistic loss, $\phi(t) = \log(1+e^{-t})$; or squared hinge loss,
$\phi(t)=\{\max(1-t,0)\}^2$.  However, other convex loss functions are
possible provided that they are differentiable, monotone, strictly
convex, and satisfy $\phi(0)=1$ \citep{bartlett:convexity06}.
As noted previously, 
\cite{Zhao:OWL12} proposed a special case of EARL called outcome weighted learning, 
 which set $\phi(t) = \max(0, 1-t)$,
$\widehat Q(\bx, a)\equiv 0$, and assumed that the propensity score
was known.  Thus, as noted previously, EARL is considerably more
general than OWL and, as shown in Section \ref{simSection}, the choice
of a non-null model for the $Q$-function and alternative surrogate
loss functions can lead to dramatically improved finite sample
performance.

\subsection{EARL via sample splitting}

To facilitate the analysis of the statistical properties of EARL, we
consider the following alternative estimator based on the sample
splitting. Let $I_1, I_2,..., I_K$ denote a random partition of the
indices $\{1, 2,...,n\}$ with $I_j\cap I_k=\emptyset$ for any
$j\neq k$ and $\cup_{k=1}^K I_k=\{1, 2,...,n\}$. We assume the size of
the partitions is comparable, i.e., $n_k=|I_k|$ with
$n_1\asymp n_2 \asymp...\asymp n_K$. In practice, $K$ is taken as a
small integer (e.g., 2, or 5) and is assumed fixed. Recall that the
EARL estimator based on the full sample is defined in
(\ref{hatFL}). In particular, the same samples are used to estimate
the nuisance functions $\hat\pi, \hat Q$ and construct the estimator
$\tilde{f}_{n}^{\lambda_{n}}$ in (\ref{hatFL}). This creates the
delicate dependence between the estimators $\hat\pi, \hat Q$ and the
samples used in the empirical risk minimization in (\ref{hatFL}). To
remove this dependence, we now modify the procedure via sample
splitting. First, for $1\leq k\leq K$, we construct estimators
$\hat\pi_k, \hat Q_k$ based on the samples in $I_k$, i.e.,
$\{(\bX_i, A_i, Y_i); i\in I_k\}$. Denote
$I_{(-k)}=\{1,...,n\}\backslash I_k$. Then, we use the remaining samples
$I_{(-k)}$ for the EARL estimator
\begin{equation}
  \widehat{f}_{n,k}^{\lambda_{nk}} = \arg\inf_{f\in\mathcal{F}}\pn^{(-k)} \left[|\widehat W_{1k}|\phi\left\{
    \mathrm{sgn}(\widehat W_{1k})f(\bX)
    \right\} + |\widehat W_{-1k}|\phi\left\{
    -\mathrm{sgn}(\widehat W_{-1k})f(\bX)
    \right\} \right]+ \lambda_{nk} \|f\|^{2},
\label{hatFLsplitk}
\end{equation}
 where $ \widehat W_{ak} = W_a(Y, \bX, A, \widehat \pi_k, \widehat{Q}_k), 
a=\in \lbrace -1,1\rbrace $ and $\pn^{(-k)}f=\frac{1}{|I_{(-k)}|}\sum_{i\in I_{(-k)}}f(X_i)$. We note that independent samples are used for estimating the nuisance functions $\pi, Q$ and the decision rule $f$. Thus, the dependence between the estimators $\hat\pi, \hat Q$ and the samples use in (\ref{hatFL}) is removed. Finally, to obtain a more stable estimator, we can aggregate the estimators
\begin{equation}
  \widehat{f}_{n}^{\lambda_{n}}= \frac{1}{K}\sum_{k=1}^K   \widehat{f}_{n,k}^{\lambda_{nk}}, 
\label{hatFLsplit}
\end{equation}
which is the final estimator based on sample splitting. While the estimator $\widehat{f}_{n}^{\lambda_{n}}$ requires more computational cost, it has important advantages over the original EARL estimator $ \tilde{f}_{n}^{\lambda_{n}}$ in (\ref{hatFL}). 
 From a theoretical perspective, one can still analyze the EARL estimator $ \tilde{f}_{n}^{\lambda_{n}}$ based on the empirical process theory. This typically requires the entropy conditions on the function classes of $\pi$ and $Q$. In comparison, we show in the following section that the sample splitting estimator $\widehat{f}_{n}^{\lambda_{n}}$ does not require this condition. To the best of our knowledge, similar sample splitting technique was first applied by \cite{bickel1982adaptive} in general semiparametric estimation problems; see also \cite{schick1986asymptotically}. Recently, this approach has received 
attention in causal inference problems as a means of relaxing technical conditions. We refer to \cite{zheng2011cross,chernozhukov2016double,robins2017minimax} for further discussion. 

\section{Theoretical properties}\label{scntheory}

%Furthermore, such results hold when
%either propensity model or outcome regression model is correct.
Let $f^*\in\mathcal{M}$ be such that
$d^*(\bx)=\sign\{f^*(\bx)\}$, and $V^* \triangleq \sup_{f\in\mathcal{M}}V(f) = V(f^*)$. Define the population risk of function $f$ as 
$$\R(f)=E(YI[A\neq\sign\{f(\bX)\}]/\pi(A;\bX)),$$   
and $\R^* \triangleq \inf_{f\in\mathcal{M}}\R(f)$. We define the risk in this way to be consistent with the convention that higher risk is less desirable; however, inspection shows that the risk equals $K - V(f)$ where $K$ is a constant that does not depend on $f$.  Thus, minimizing risk is equivalent to maximizing value, and $V^*-V(f)=\R(f)-\R^*$. Accordingly, for a convex function $\phi$, we define the $\phi$-risk
$$\R^m_\phi(f)=E [|{W}^m_{1}|\phi\left\{\mathrm{sgn}({W}^m_{1})f(\bX)
\right\} + |{W}^m_{-1}|\phi\left\{-\mathrm{sgn}({W}^m_{-1})f(\bX)
\right\}].$$
By construction, $\R^m_\phi(f)$ is convex; we assume
that it has a unique minimizer  and that
$\R_{\phi}^{m*} \triangleq \inf_{f\in\mathcal{M}}\R_{\phi}^m(f)$.  The
following result is proved in Web Appendix C.

\begin{prop}\label{thm_fisher}
Assume that either $\pi^m(a;\bx) = \pi(a;\bx)$ or $Q^m(\bx,a)=Q(\bx,a)$.
Define $\tilde{f} =\argmin_{f\in\mathcal{M}}R_\phi^m(f)$ and 
$c_m(\bx)=E\{|W_1(Y, \bx, A, \pi^m, Q^m)|+|W_{-1}(Y, \bx, A, \pi^m, Q^m)|\}$.
Then: 
\begin{itemize}
\item[(a)]  $\D^*(\bx)=\sign\{\tilde f(\bx)\}$;
\item[(b)] and
   $$
   \psi\left\{\frac{V^*-V(f)}{\sup_{\bx\in \re^p}
       c_m(\bx)}\right\} \leq \frac{\R_{\phi}^m(f) -
     \R_{\phi}^{m*}}{\inf_{\bx\in \re^p }c_m(\bx)},
   $$
   where  $\psi(\theta)=|\theta|$ for hinge loss,
   $\psi(\theta)=1-\sqrt{1-\theta^2}$ for exponential loss,
   $\psi(\theta)=(1+\theta)\log(1+\theta)/2+(1-\theta)\log(1-\theta)/2$
   for logistic loss, and $\psi(\theta)=\theta^2$ for squared hinge
   loss.
\end{itemize}
\end{prop} 
\noindent
Part (a) of the preceding proposition states that if
  either the model for the propensity score or for the $Q$-function is
  correctly specified, then the EARL procedure, optimized over the
  space of measurable functions, is Fisher consistent for the optimal
  rule.  Part (b) bounds the difference between $V(f)$ and $V^*$
through the surrogate risk difference
$\R_{\phi}^m(f) -\R_{\phi}^{m*}$.   
The different forms of $\psi(\cdot)$ are due to the fact that
different loss functions 
induce different distance measures
of closeness of $f(x)$ to the true $f^*(x)$.  We use these risk bounds to derive
bounds on the convergence rates of the value of EARL estimators constructd
using sample splitting.

%\subsection{Convergence rates of the EARL estimator} 
Let  $\Pi$ denote the function spaces to which
the postulated models for  $\pi(a;\bx)$ 
belong; i.e., the estimator $\widehat {\pi}(a;\bx)$ belongs to $\Pi$.
Similarly, let $\mathcal{Q}$ denote a postulated class of models
for $Q(\bx, a)$.    
In this section, we allow the approximation
space, $\mathcal{F}$, to be 
arbitrary subject to complexity constraints; our results allow
both parametric or non-parametric classes of models.
Our primary result is a bound on the rate of convergence of
$V^*-V(\widehat{f}_{n}^{\lambda_{n}})$ in terms of the $\phi$-risk
difference
$\R_{\phi}^m(\widehat f_n^{\lambda_n}) - \R_{\phi}^{m*}$.  

For any
$\epsilon>0$ and measure $P$, let $N\left\lbrace 
\epsilon, \mathcal{F}, L_2(P)\right\rbrace$
denote the covering number of the space $\mathcal{F}$, i.e.,
$N\left\lbrace \epsilon,\mathcal{F},L_2(P)\right\rbrace$ 
is the minimal number of closed
$L_2(P)$-balls of radius $\epsilon$ required to cover $\mathcal{F}$
\citep{kosorok:ep08}.  Denote $\|f\|_{P,2}^2=Ef^2(\bX)$. We
make the following assumptions.
\begin{assumption}
  {There exists $M_Q > 0$ such that $|Y|\le M_{\mathcal{Q}}$ and $|Q(\bx, a)| \le  M_{\mathcal{Q}}$ for all 
    $(\bx, a)\in \re^p \times \{-1,1\}$ and
    $Q\in\mathcal{Q}$; there exists $0 < L_{\Pi} < M_{\Pi} < 1$
    such
    that $L_{\Pi} \le \pi(a;\bx) \le M_{\Pi}$ for all $(\bx, a)\in
   \re^p \times \{-1,1\}$ and $\pi \in \Pi$.}
\label{bdd_assp}
\end{assumption}
 
\begin{assumption}
  There exist constants $0 <  v < 2$ and
  $c< \infty$ such that for all $0 < \epsilon \le 1$:
$ \sup_P \log N\left\lbrace \epsilon,
   \mathcal{F}, L_2(P)\right\rbrace \leq c {\epsilon}^{-v}, $  where the supremum is taken over all
  finitely discrete probability measures $P$.  
\label{coveringnumber_assp} 
\end{assumption} 
 
\begin{assumption}
For some $\alpha$, $\beta>0$, $E\|\widehat
\pi_k(a;\bx)-\pi(a;\bx)\|^2_{P,2}=O(n^{-2\alpha})$ and $E\|\widehat
Q_k(\bx,a)-Q(\bx,a)\|^2_{P,2}=O(n^{-2\beta})$ for $a=\pm 1$ and $1\leq k\leq K$. 
\label{Wrate}
\end{assumption}  

%\begin{assumption}
%The functional space $\mathcal{F}$ is constructed using linear combinations from a fixed base class $\mathcal{H}$, written  $\mathcal{F} = \mathrm{absconv}(\mathcal{H})$, where $\mathcal{H}$ has finite Vapnik-Chervonenkis (VC) dimension $v$. 
%The covering number of the functional space $\mathcal{F}$ is bounded as follows,
%  $$ \sup_P \log N(\epsilon,
%   \mathcal{F}, L_2(P)) \leq C_v{\epsilon}^{-v}, $$ 
% where $0< v\leq 2$ and $C_v$ is a constant depending on $v$. 
% \label{coveringF_assp}
%\end{assumption}
\noindent
Assumption \ref{bdd_assp} assumes outcomes are bounded, which often holds in practice. Otherwise, we can always use a large constant to bound the outcome. We also assume propensity scores are bounded away from 0 and 1, which is a standard condition for the identification of the treatment effect in causal inference. Assumption \ref{coveringnumber_assp} controls the complexity of the
function spaces for estimating an optimal ITR. For example,
if $\mathcal{F}$ is composed of  
linear combinations of elements in a fixed base class, $\mathcal{H}$,
where 
$\mathcal{H}$ has
finite Vapnik-Chervonenkis (VC) dimension $vc$, then there exists a
constant $c_{vc}$, depending on $vc$, so that
$ \sup_P \log N\left\lbrace \epsilon, \mathcal{F}, L_2(P)\right\rbrace \leq
c_{vc}{\epsilon}^{-2vc/(vc+2)}$
(Theorem 9.4, \citet{kosorok:ep08}).  We note that the entropy conditions on $\mathcal{Q}$ and $\Pi$ are not needed by using the sample splitting technique, due to  the independence between estimating $\pi, Q$ and estimating $f$.

Assumption \ref{Wrate} specifies the rate of convergence of the
estimators $\widehat\pi$ and $\widehat Q$ in terms of the $\|\cdot\|_{P,2}$ norm. It
is well known that the $L_2$ rate of convergence is related to the
smoothness of the function classes $\mathcal{Q}$ and $\Pi$ and the
dimension of $\bX$. For instance, if $\mathcal{Q}$ corresponds to the
Holder class with smoothness parameter $s$ on the domain $[0,1]^p$,
then Theorem 7 of \cite{newey1997convergence} implies $E\|\widehat
Q(\bx,a)-Q(\bx,a)\|^2_{P,2}=O_p(K/n+K^{-2s/p})$,
where $\widehat Q(\bx,a)$ is the regression spline estimator and $K$
is the number of basis functions.

%The difference in  $\phi$-risk $\R_{\phi}^m(f) -\R_{\phi}^{m*}$ can be split into an estimation error term and an approximation error term. 
Define the approximation error incurred by
optimizing over $\mathcal{F}$ as
\begin{equation}
\mathcal{A}(\lambda_n)= \inf_{f\in \mathcal{F}} \Big(\lambda_n\|f\|^2+
\sum_{a = \pm 1}E\left[ {W_a^m\phi\{a \cdot f(\bX)\}}\right]\Big)-\inf_{f\in
  \mathcal{M}}\sum_{a=\pm 1 }E\left[ {W_a^m\phi\{a \cdot {f}(\bX)\}}\right].
\label{approxfcn}  
\end{equation}
The following result on the risk bound is the main result in this section and is proved in the Web Appendix D.

\begin{thm}\label{thm_rate}
  Suppose that assumptions 1-4 hold, $\lambda_n \rightarrow 0$.  Define $c_m(\bx)=E\{|W_1^{m}|\cond\bX=\bx,A =1\}+
E\{|W_{-1}^m|\cond\bX=\bx,A =-1\}$. If $Q^m(\bx,a)=Q(\bx,a)$ and $\pi^m(a;\bx) = \pi(a;\bx)$, then 
\begin{align*}
  \psi\left\{\frac{V^* -V(\widehat{f}_{n}^{\lambda_{n}})}{\sup_{\bx\in\re^p}c_m(\bx)}\right\} & \lesssim    \frac{1}{\inf_{\bx\in\re^p} c_m(\bx)}\cdot\Big[\mathcal{A}(\lambda_n)+n^{-\frac{2}{v+2}}\lambda_n^{-\frac{v }{v +2}}+ n^{-1}\lambda_n^{-1}
\\
&~~~~+\lambda_n^{-1/2}n^{-(\alpha+\beta)}+\lambda_n^{-1/2} (n^{-(1/2+\alpha)}+n^{-(1/2+\beta)}) 
      \Big].
\end{align*} 
\end{thm}

In all cases considered, the function $\psi$ 
is invertible on $[0,1]$, and its inverse is monotone non-decreasing.
Thus, for sufficiently large $n$ 
(making the right-hand-side of the equation sufficiently small)
the inequality can be re-arranged to yield a bound on 
$V^*-V(\widehat{f}_{n}^{\lambda_n})$.  The 
form of $\psi^{-1}$ dictates the tightness of the bound as a function
of the $\phi$-risk.  According to Lemma 3 in Bartlett et al (2006), a flatter loss function leads to better bound on $\psi$ function. In other words, a flatter loss function gives better bounds on $V^* - V(f)$ in terms of $\mathcal R_\phi^m(f) - \mathcal R_\phi^{m*}$.  In this respect, hinge-loss can be seen to
provide the tightest bound; however, the $\phi$-risk is not directly
comparable across different loss functions as they are not on the same
scale. % In application, one might select a loss function by choosing the one with highest cross-validated value \citep{van2007super}.
 
The right hand side of the bound in Theorem \ref{thm_rate}
consists of three parts: the
approximation error $\mathcal{A}(\lambda_n)$ due to the size of the
approximation space $\mathcal{F}$, the error $n^{-\frac{2}{v
    +2}}\lambda_n^{-\frac{v}{v +2}}+ n^{-1}\lambda_n^{-1}$ due to the
estimation in the function space $\mathcal{F}$,
 and 
the error
$\lambda_n^{-1/2}n^{-(\alpha+\beta)}+\lambda_n^{-1/2} (n^{-(1/2+\alpha)}+n^{-(1/2+\beta)})$ incurred from plugging the estimators $\widehat\pi_k$ and $\widehat Q_k$. As
expected, the approximation error decreases as the complexity of the
class $\mathcal{F}$ increases, whereas the estimation error increases
with the complexity of the class $\mathcal{F}$ and decreases as the
sample size increases. %In general, the derived rate depends on the complexity of $\mathcal{Q}$, $\Pi$, and $\mathcal{F}$.  The max operator illustrates that the rate of convergence for the value is dictated, in part, by the most complex of these three classes.  

For the error incurred from plugging the estimators $\widehat\pi_k$ and
$\widehat Q_k$, the component $\lambda_n^{-1/2} (n^{-(1/2+\alpha)}+n^{-(1/2+\beta)})$ converges to 0 
faster than $\lambda_n^{-1/2}n^{-(\alpha+\beta)}$ in regular statistical models (i.e., $\alpha,\beta\leq 1/2$). Thus, it suffices to only look at the term $\lambda_n^{-1/2}n^{-(\alpha+\beta)}$. This term  can shrink to 0 sufficiently
fast as long as one of the estimators $\widehat\pi_k$ and $\widehat Q_k$
has a fast rate,  due to the multiplicative form of the estimation
error. For example, if $\alpha = \beta = 1/4$, the error from plugging
the estimators $\widehat\pi$ and $\widehat Q$ is
$n^{-1/2}\lambda_n^{-1/2}$. Hence, the rate of the proposed method is
faster compared with the outcome weighted learning method, which is
developed based on an IPWE and does not enjoy this multiplicative form
of the errors. 
This phenomenon can be viewed as a nonparametric version of the double
robustness property \cite[see][for additional discussion]{fan2016improving,benkeser2017doubly}.  Compared with the results in \citet{athey2017efficient}, we allow for the surrogate loss to replace the 0-1 loss in solving for the optimizer. While the orders in the bound of convergence rates are comparable, the differences in the constants in the bounds might be due to the application of the surrogate function.

% the error incurred from plugging the estimators $\widehat\pi$ and $\widehat Q$ is dominated by $\lambda_n^{-1/2}n^{-(\alpha+\beta)}$, which depends on the product of the estimation errors of $\widehat\pi$ and $\widehat Q$.   

\begin{remark}
If $\alpha=\beta$ and 
\begin{equation}\label{eqthm11}
n^{2\alpha-1}\lambda_n^{-1/2}\rightarrow\infty, ~\textrm{or}~n^{2\alpha(v+2)-2}\lambda_n^{1-v/2}\rightarrow\infty,
\end{equation} 
then 
\begin{equation}\label{eqthm12}
  \psi\left\{\frac{V^* -V(\widehat{f}_{n}^{\lambda_{n}})}{\sup_{\bx\in\re^p}c_m(\bx)}\right\}  \lesssim    \frac{1}{\inf_{\bx\in\re^p} c_m(\bx)}\cdot\Big[\mathcal{A}(\lambda_n)+n^{-\frac{2}{v +2}}\lambda_n^{-\frac{v }{v +2}}+ n^{-1}\lambda_n^{-1}
      \Big],
\end{equation}
where the upper bound is of the same order as that obtained if the conditional mean $Q(\bx,a)$ and propensity score $\pi(a;\bx)$ are known. We note that the additional constraints on $\alpha$ and $v$ in (\ref{eqthm11}) are necessary to obtain the fast rate of convergence (\ref{eqthm12}). For instance, if the function classes $\mathcal{Q}$ and $\Pi$ are indexed by finite dimensional parameters, we can obtain $\alpha=\beta=1/2$ under mild conditions. As a result, the first condition in (\ref{eqthm11}) holds and the fast rate of convergence (\ref{eqthm12}) is applied. On the other hand, if  $\mathcal{F}$ is a simple class but $\|\widehat \pi - \pi\|_{P,2}$ and $\|\widehat Q - Q\|_{P,2}$ converge at slower rates, the rate for $V^*-V(\widehat{f}_{n}^{\lambda_n})$ will be driven by $\lambda_n^{-1/2}n^{-(\alpha+\beta)}$. 
\end{remark}

To estimate the value of the optimal treatment rule $V^*$, one can aggregate the empirical value of the sample splitting estimator $\hat f_{n,k}^{\lambda_{n,k}}$ in each subsamples $I_{(-k)}$, i.e., $\bar V=\frac{1}{K}\sum_{k=1}^K \hat V_{(-k)}(\hat f_{n,k}^{\lambda_{n,k}})$, where 
$$
\hat V_{(-k)}(f)= \pn^{(-k)} \left[|\widehat W_{1k}|\phi\left\{
    \mathrm{sgn}(\widehat W_{1k})f(\bX)
    \right\} + |\widehat W_{-1k}|\phi\left\{
    -\mathrm{sgn}(\widehat W_{-1k})f(\bX)
    \right\} \right],
$$
The following corollary, provides a corresponding bound on the rate for $V^* - \bar V$. The proof is given in Web Appendix D. 
\begin{cor}
Suppose that assumptions 1-4 hold, and $\lambda_n \rightarrow 0$.  If $Q^m(\bx,a)=Q(\bx,a)$ and $\pi^m(a;\bx) = \pi(a;\bx)$, then 
\begin{align*}
V^* - \bar V  \lesssim \frac{1}{K}\sum_{k=1}^K[V^* -V(\hat f_{n,k}^{\lambda_{n,k}})] +n^{-1/2}+\lambda_n^{-1/2}n^{-(\alpha+\beta)}+\lambda_n^{-1/2} (n^{-(1/2+\alpha)}+n^{-(1/2+\beta)}), 
\end{align*} 
where 
\begin{align*}
\frac{1}{K}\sum_{k=1}^K[V^* -V(\hat f_{n,k}^{\lambda_{n,k}})] &\lesssim \sup_{\bx\in\re^p}c_m(\bx) \psi^{-1}\Big[ \frac{1}{\inf_{\bx\in\re^p} c_m(\bx)}\cdot\Big\{\mathcal{A}(\lambda_n)+n^{-\frac{2}{v+2}}\lambda_n^{-\frac{v }{v +2}}+ n^{-1}\lambda_n^{-1}\\
&~~~~+\lambda_n^{-1/2}n^{-(\alpha+\beta)}+\lambda_n^{-1/2} (n^{-(1/2+\alpha)}+n^{-(1/2+\beta)}) \Big\}\Big].
\end{align*} 
\end{cor}

\begin{remark}
  \cite{athey2017efficient} and \cite{kitagawa2017should} investigated
  the binary-action policy learning problem, and established a risk
  bound of $n^{-1/2}$ for both known propensities
  \citep{kitagawa2017should} and unknown propensities
  \citep{athey2017efficient}.  However, they considered a restricted
  class of decision rules and subsequent risk bound were
  established with respect to the optimal rule within this restricted
  class. Hence, there was not consideration of the approximation
  error. In contrast, we considered the optimal rule within the space
  consisting of all measurable functions from $\mathbb{R}^p$ (the
  covariate space) to $\{-1,1\}$ (the treatment space). We used a
  smaller space, for example, a reproducing kernel Hilbert space, to
  approximate the policy space and to avoid overfitting. This led to a
  tradeoff between approximation and estimation error, and $\lambda_n$
  was a tuning parameter to control this bias-variance
  tradeoff. Consequently, the achieved convergence rates are different.
\end{remark}
\section{Simulation experiments}\label{simSection}
We compare EARL estimators with: $Q$-learning fit using ordinary least
squares \citep[QL,][]{qian:itr11}; estimating the optimal rule within
a restricted class based on an AIPW estimator
\citep[AIPWE,][]{zhang2012robust}; and outcome weighted learning
\citep[OWL,][]{Zhao:OWL12}. Comparisons are made in terms of the
average value of the rule estimated by each method. For $Q$-learning,
we fit a linear model for the $Q$-function that includes all two-way
interactions between predictors and pairwise interactions between
these terms and treatment. In the AIPWE method, an AIPW estimator for
the value function is constructed and then the optimal linear rule
that maximizes the AIPW estimator is identified via a genetic
algorithm. Similar to EARL, both a propensity score model and a
regression model need to be fitted in AIPWE. We will use the same set
of models in EARL and the AIPWE, which are detailed in below.
%will use an augmented estimator, which depends on both a propensity
%score model and a regression model, to construct the weights. 
For OWL, we use a linear decision rule; recall that OWL is a special
case of EARL with $\widehat Q(\bx, a)\equiv 0$,
$\phi(t) = \max(0, t)$, and a known propensity score.  All estimation
methods under consideration require penalization;
we choose the amount of penalization using 10-fold cross-validation of
the value.  Within the class of EARL estimators, we considered hinge,
squared-hinge, logistic, and exponential convex surrogates.  
An implementation of EARL is available in the R package `DynTxRegime;'
this package also includes implementations of AIPWE and OWL and
therefore
can be used to replicate the simulation studies presented here. We included an example for implementing EARL method using `DynTxRegime' package in Web Appendix H. 
%We use this package to implement AIPWE and OWL as well.  
%Additional modeling choices for EARL estimators are given below, in order to designate these models as correctly or incorrectly specified, we first describe the class of generative models used in our simulation experiments.

We consider generative models of the form:
$\bX=(X_1, \ldots, X_p) \sim_{i.i.d.}N (0,1)$ with $p=10$; treatments
are binary, taking the values in $\{-1,1\}$ according to the model
$ p(A=1\cond \bX) = \exp\{\ell(\bX)\}/[1+\exp\{\ell(\bX)\}],$ where
$\ell(\bx) = x_1 + x_2 + x_1x_2$ in Scenario 1, and $\ell(\bx) = 0.5x_1 - 0.5$ in Scenario 2; $ Y = \sum_{j=1}^p X_j^2 + \sum_{j=1}^p X_j+Ac(\bX)+\epsilon, $
where $\epsilon \sim N(0,1)$, and $c(\bx)= x_1+x_2 -0.1$. 
%\subsection{Double robustness}
Write $\bX^2$ to denote $(X_1^2,\ldots,X_p^2)$. The following modeling
choices are considered for the propensity and outcome regression
models.
\begin{enumerate} 
\item[CC.] A correctly specified logistic regression model for
  $\pi(A; \bX)$ with predictors $X_1$, $X_2$ and $X_1X_2$ in Scenario 1, and with predictor $X_1$ in Scenario 2; and a 
correctly specified linear
  regression model for $Q(\bX,A)$ with predictors $\bX,\bX^2,A, X_1A$ and
  $X_2A$ in both scenarios.
\item[CI.] A correctly specified logistic regression model for $\pi(A; \bX)$ 
  with predictors $X_1$, $X_2$ and $X_1X_2$ in Scenario 1, and with predictor $X_1$ in Scenario 2; and an incorrectly
  specified linear model for
  $Q(\bX,A)$ with predictors
  $\bX, A, \bX A$ in both scenarios.
\item[IC.] An incorrectly specified logistic regression model for
  $\pi(A; \bX)$ with predictors $\bX$ in Scenario 1, and without any
  predictors in Scenario 2; and a correctly specified linear model for
  $Q(\bX,A)$ with predictors $\bX,\bX^2,A, X_1A$ and $X_2A$ in both
  scenarios.
\item [II.] An incorrectly specified logistic regression model for
  $\pi(A; \bX)$ with predictors $\bX$ in Scenario 1, and without any predictors in Scenario 2;  and an incorrectly specified linear model for
  $Q(\bX,A)$  with predictors $\bX, A, \bX A$ in both scenarios.
\end{enumerate}

We use the same model specifications to carry out AIPWE, and denote them as CC-A, CI-A, IC-A, and II-A correspondingly. 
For the OWL method, we use correct and incorrect propensity models to construct the ITRs, and denote them as C. and I. respectively. Similarly, we use Q-learning to construct the ITRs based on correct and incorrect regression models, and term them as .C and .I respectively. 

%Because OWL does not depend on an outcome model, we pool the performance for OWL across cases $CC$ and $CI$ into a single setting denoted `$C\cdot,$' and pool performance for OWL across $IC$ and $II$ into a single setting denoted `$I\cdot'$.  Similarly, because QL does not depend on the propensity model so we pool the performance across $CC$ and $IC$ into a single setting denoted `$\cdot C$' and $CI$ and $II$ into a single setting denoted `$\cdot I$'. We use correct models for both the propensity score and the outcome regression to construct the weights  used in weighted CART, which  is denoted  $CC-C$.

% We also conducted a set of comparisons within the
% EARL, for example, different surrogate loss functions, different
% working models for the outcomes and propensity scores for the
% augmented procedure and different samples sizes. To avoid overfitting,
% we generated a large testing dataset of sample size 10000 to evaluate
% the estimated regimes.  Specifically, we constructed the regimes using
% different methods with the training datasets, and then obtained the
% outcomes due to the estimated regimes for the subjects in the testing
% dataset. Expected values were subsequently calculated by averaging
% over 10000 subjects.

\begin{figure}[h!]
\caption{Boxplots for Scenario 1 results under QL, AIPWE, and OWL and EARL using logistic loss.}
\label{figIPWvsAIPW1}
\begin{center}
  \includegraphics[width=2.8in,height=2.8in]{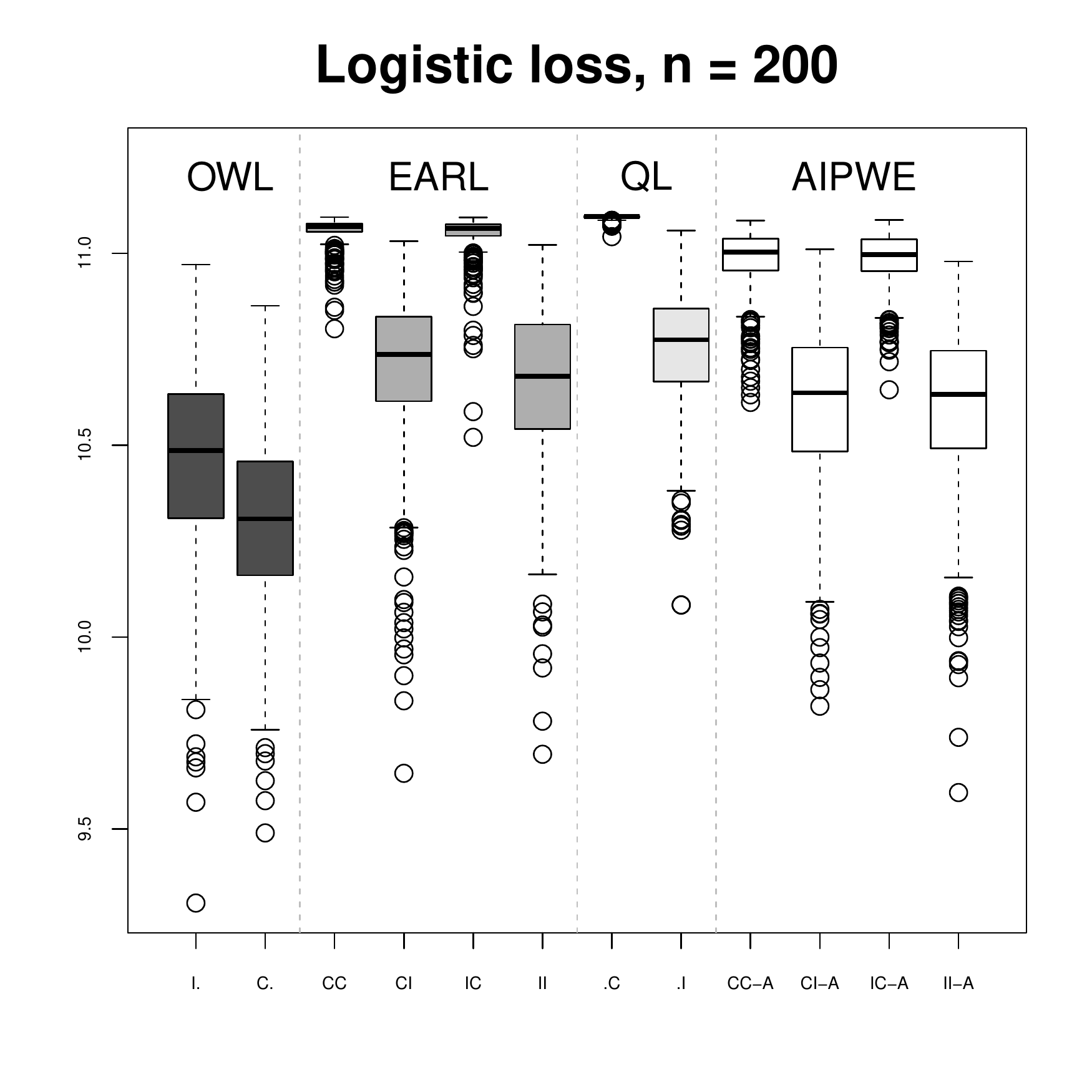}
\includegraphics[width=2.8in,height=2.8in]{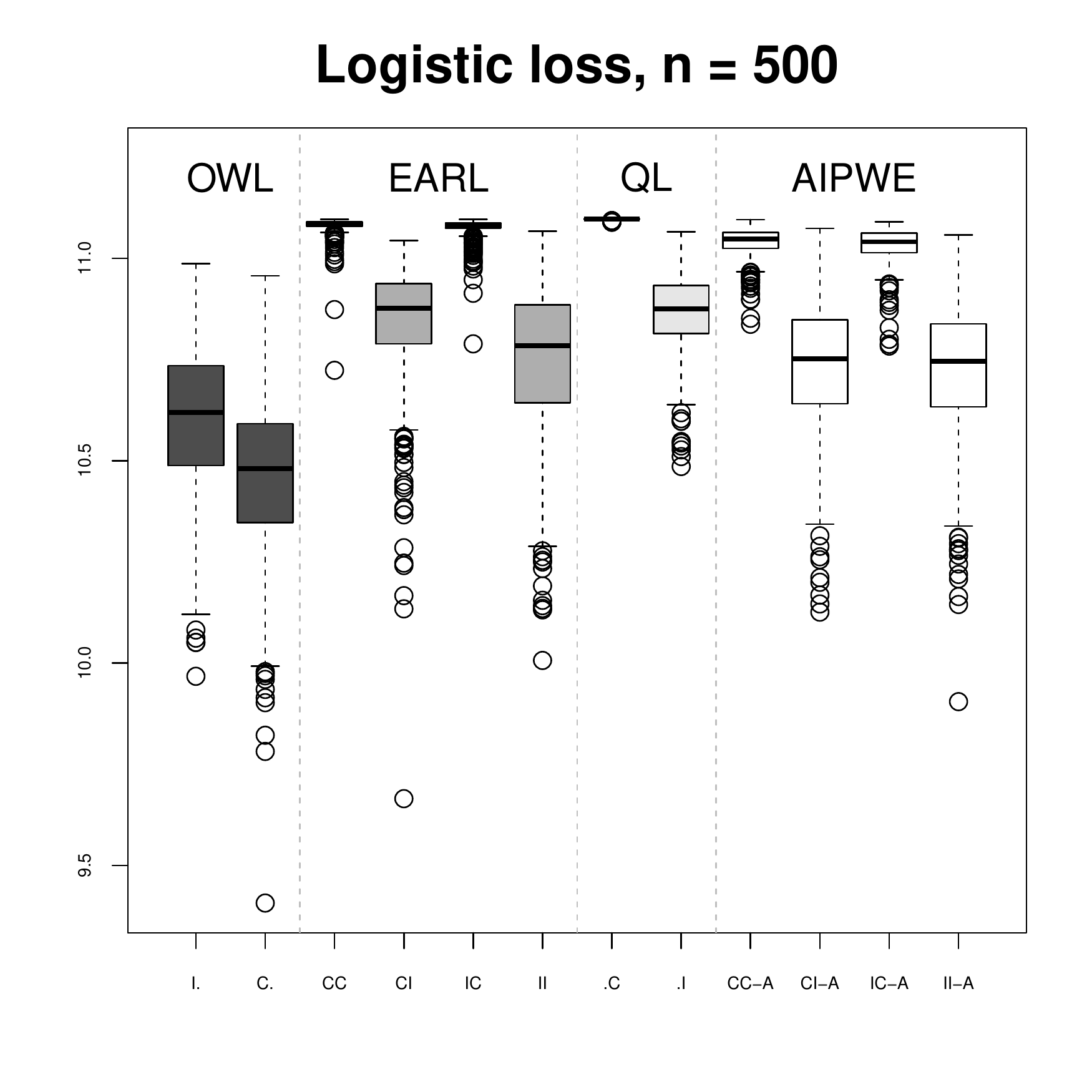}
\includegraphics[width=2.8in,height=2.8in]{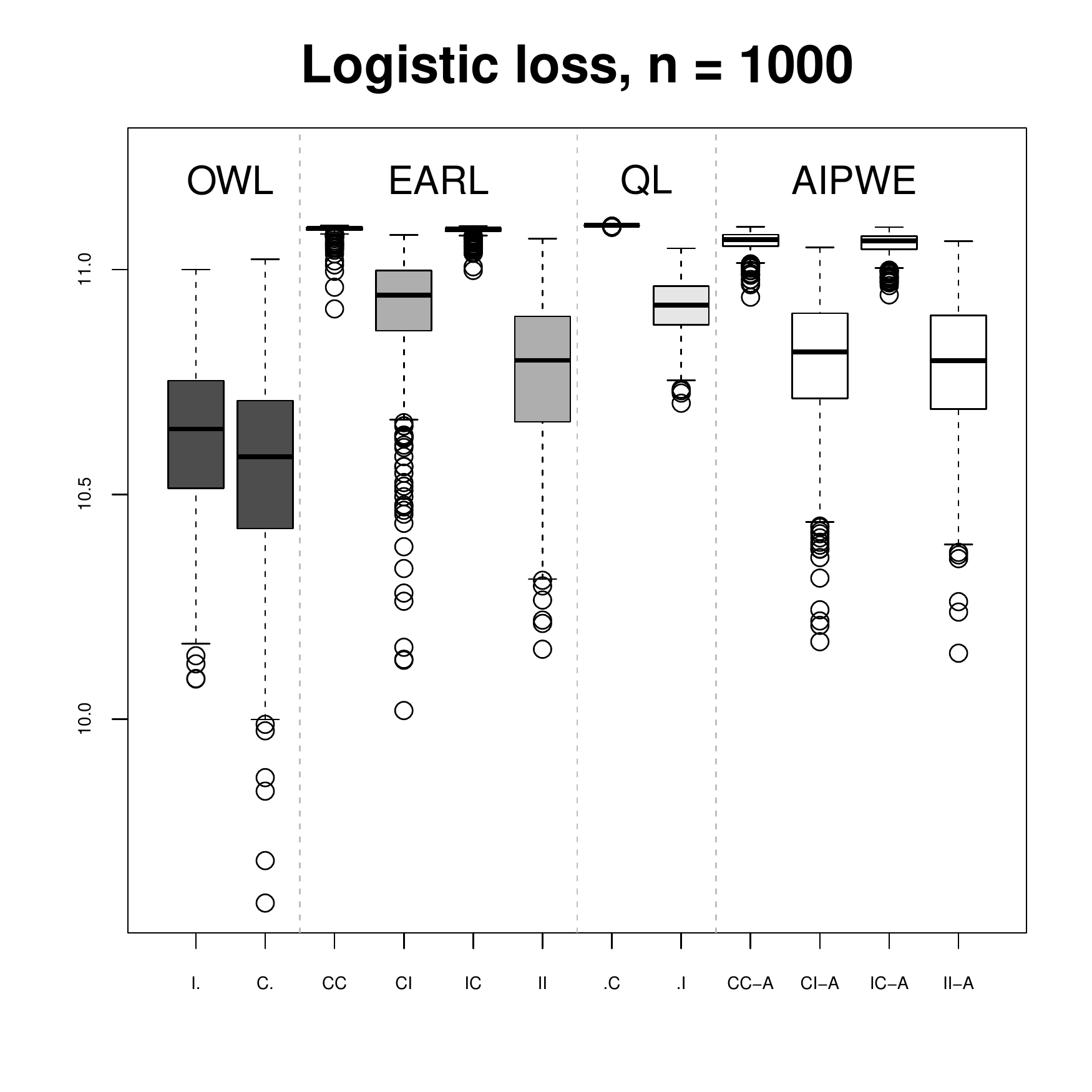}
\includegraphics[width=2.8in,height=2.8in]{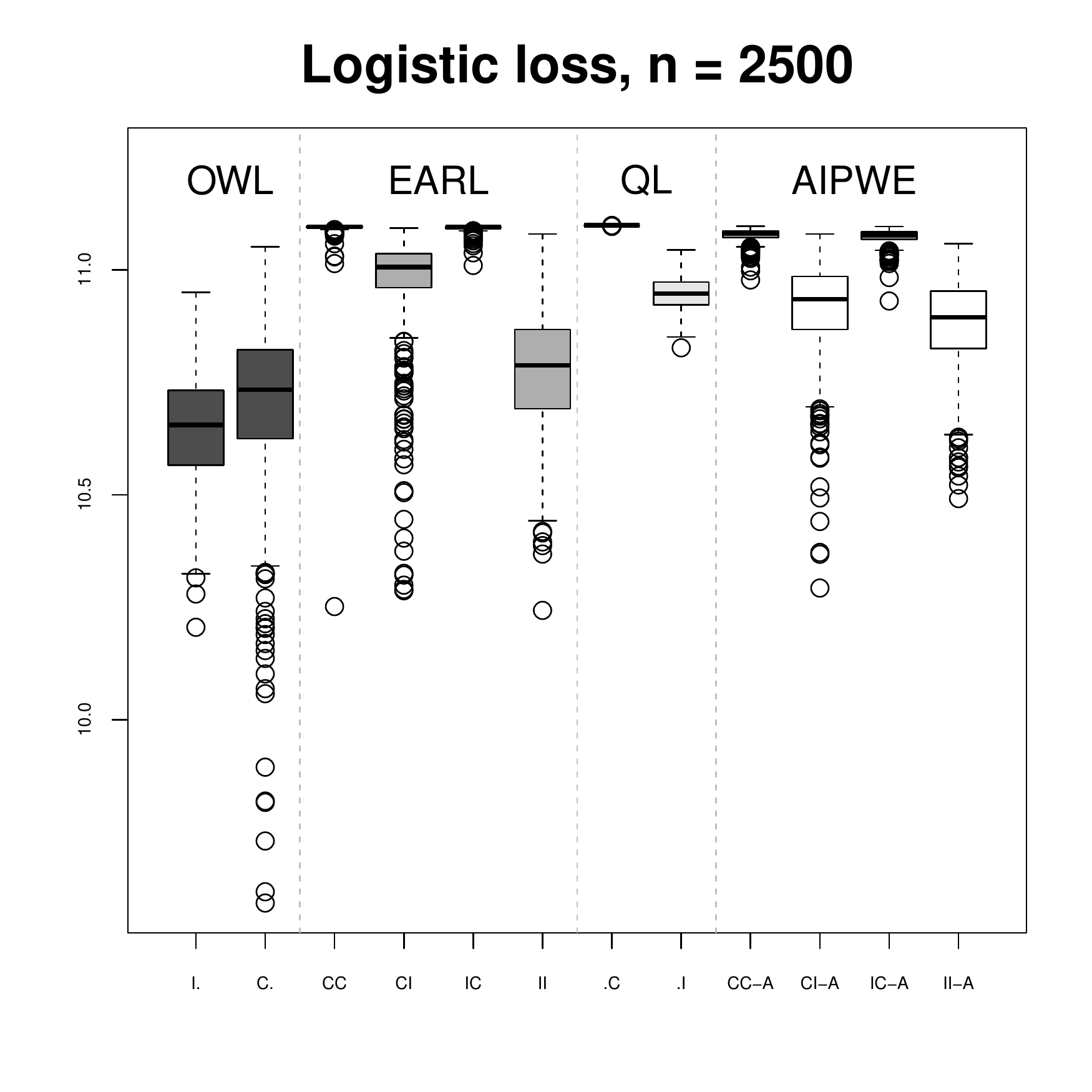}
\includegraphics[width=2.8in,height=2.8in]{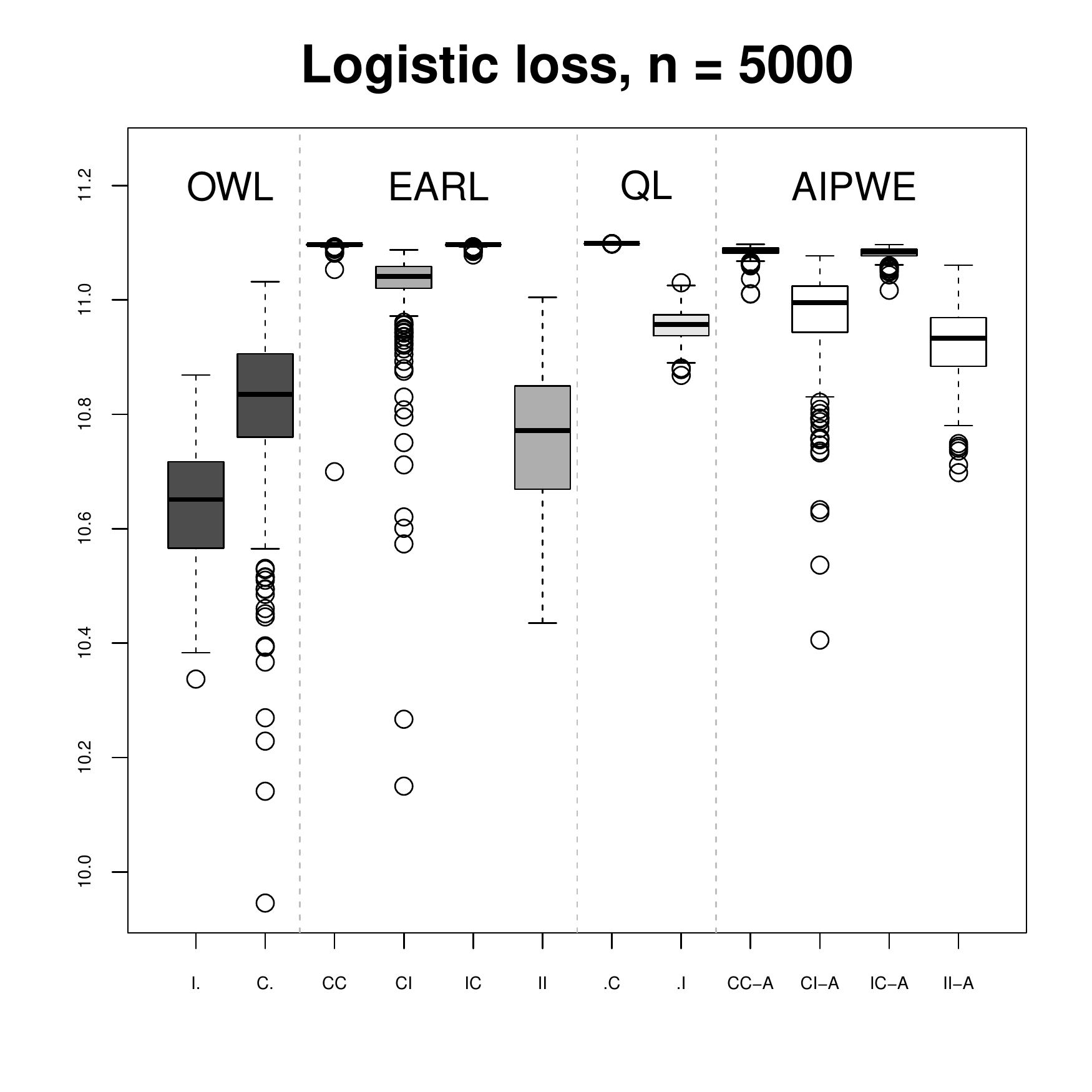}
\includegraphics[width=2.8in,height=2.8in]{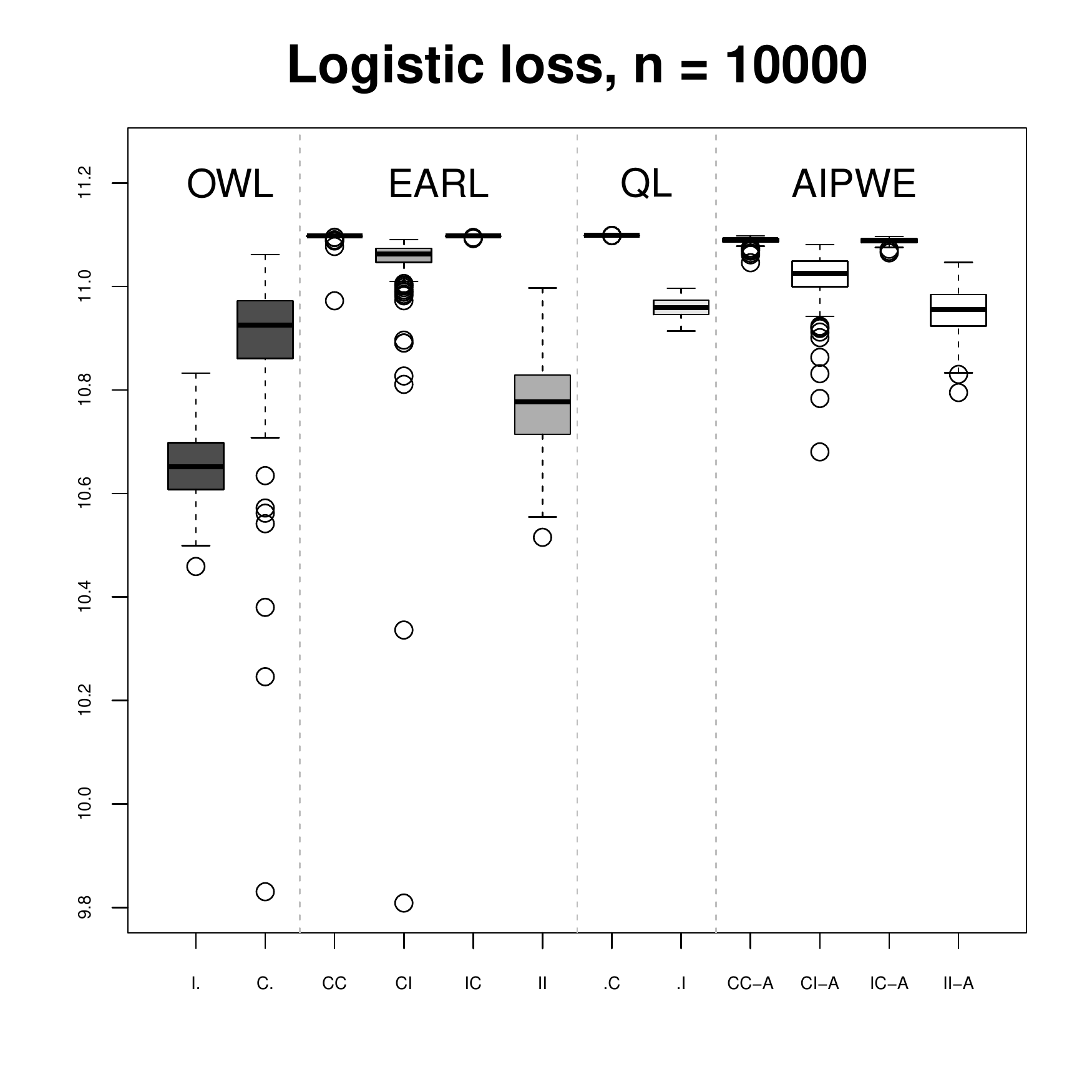}
\end{center}
\end{figure}

We consider sample sizes 200, 500, 1000, 2500, 5000 and 10000. We generate a large
validation data set (size 10000) and 500 training sets under each
sample size.  The ITRs are constructed based on one training set out
of 500 replicates using competing methods. For implementing EARL, we
use logistic loss. We observe similar patterns for other surrogate
loss functions (see the Web Appendix). We carry out cross-validation to
select $\lambda_n$ among a pre-specified set of values
$(2^{-5,}, 2^{-4},\ldots, 2^5)$. 
%The selected $\lambda_n$ maximized the
%average of the estimated values on the validation data.  
Then we
calculate the mean response had the whole population followed the rule
(the value function) by averaging the outcomes over 10000 subjects
under the estimated ITRs in the validation data set.  Thus, there are 500
values of the estimated rules on the validation set for each sample
size.  Boxplots of these values are shown in  Figures
\ref{figIPWvsAIPW1} and \ref{figIPWvsAIPW2}.  The performance of OWL
was generally worse than that of the EARL estimator or QL.  The AIPWE
method exhibits a larger bias and a higher variance compared to the
proposed method, while running approximately 200 times slower.  As expected, the QL method works best when the model is correctly
specified but can perform poorly when this model is misspecified.  
 
 \begin{figure}[h!]
\caption{Boxplots for Scenario 2 results under QL, AIPWE, and OWL and EARL using logistic loss.}
\label{figIPWvsAIPW2}
\begin{center}
 
  \includegraphics[width=2.8in,height=2.8in]{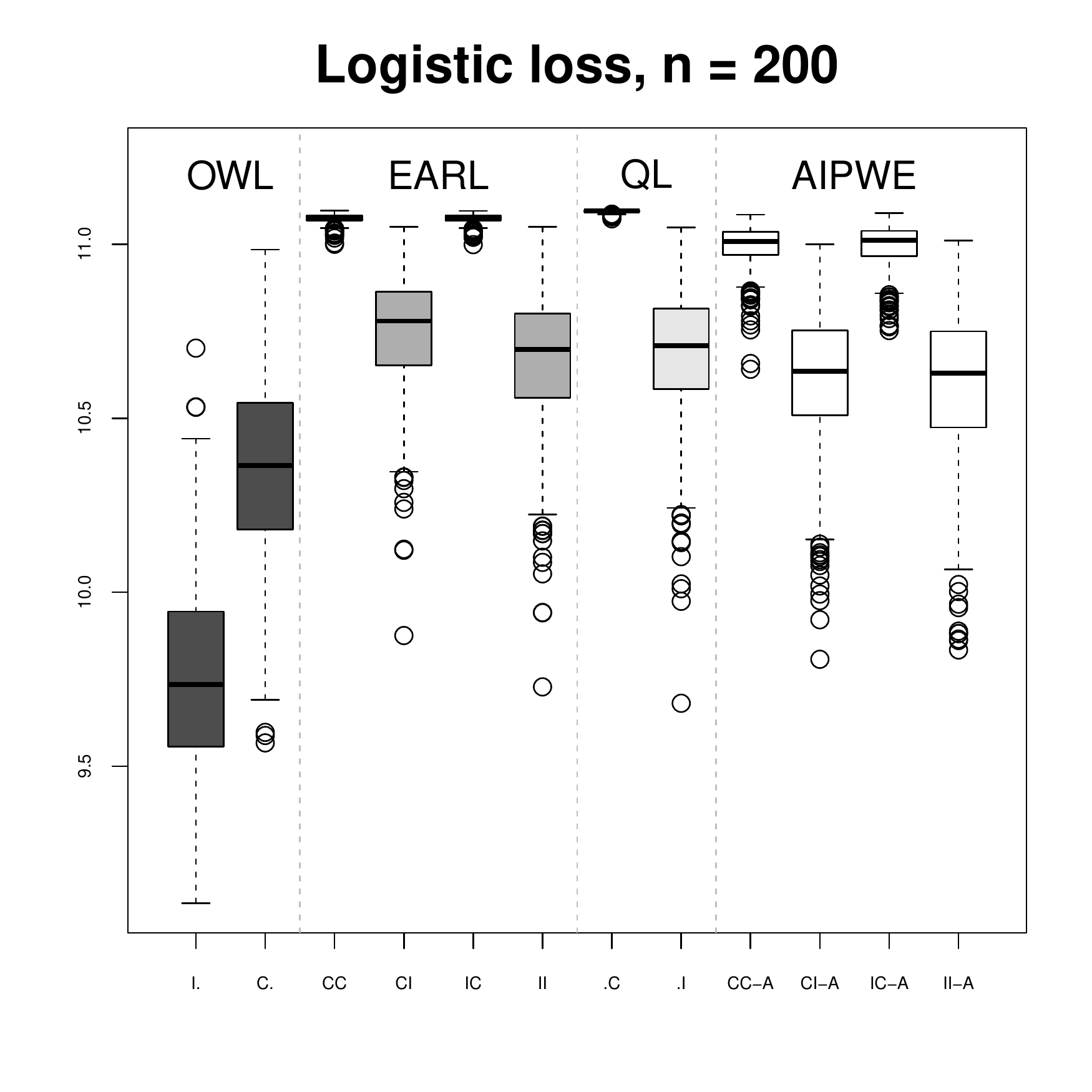}
\includegraphics[width=2.8in,height=2.8in]{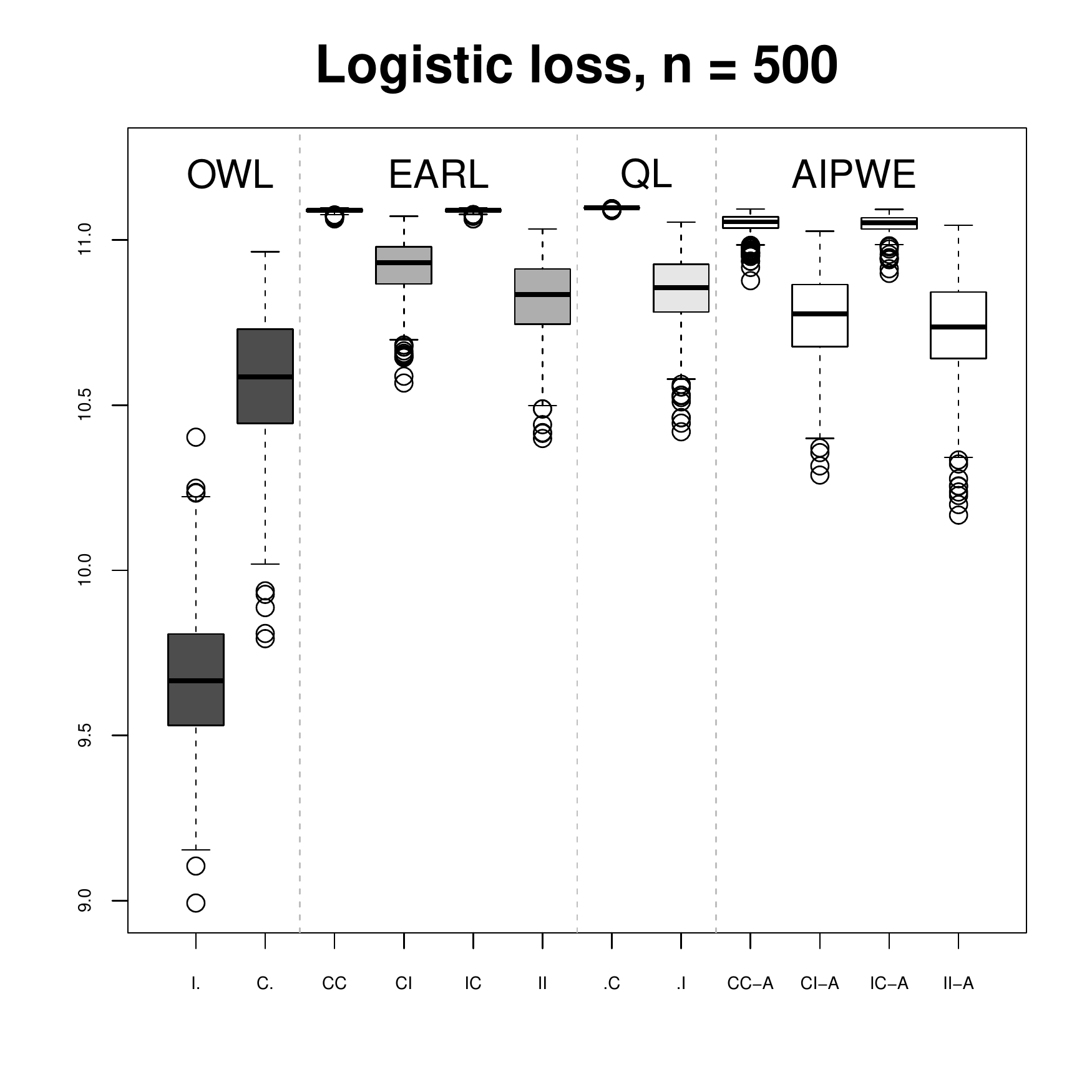}
\includegraphics[width=2.8in,height=2.8in]{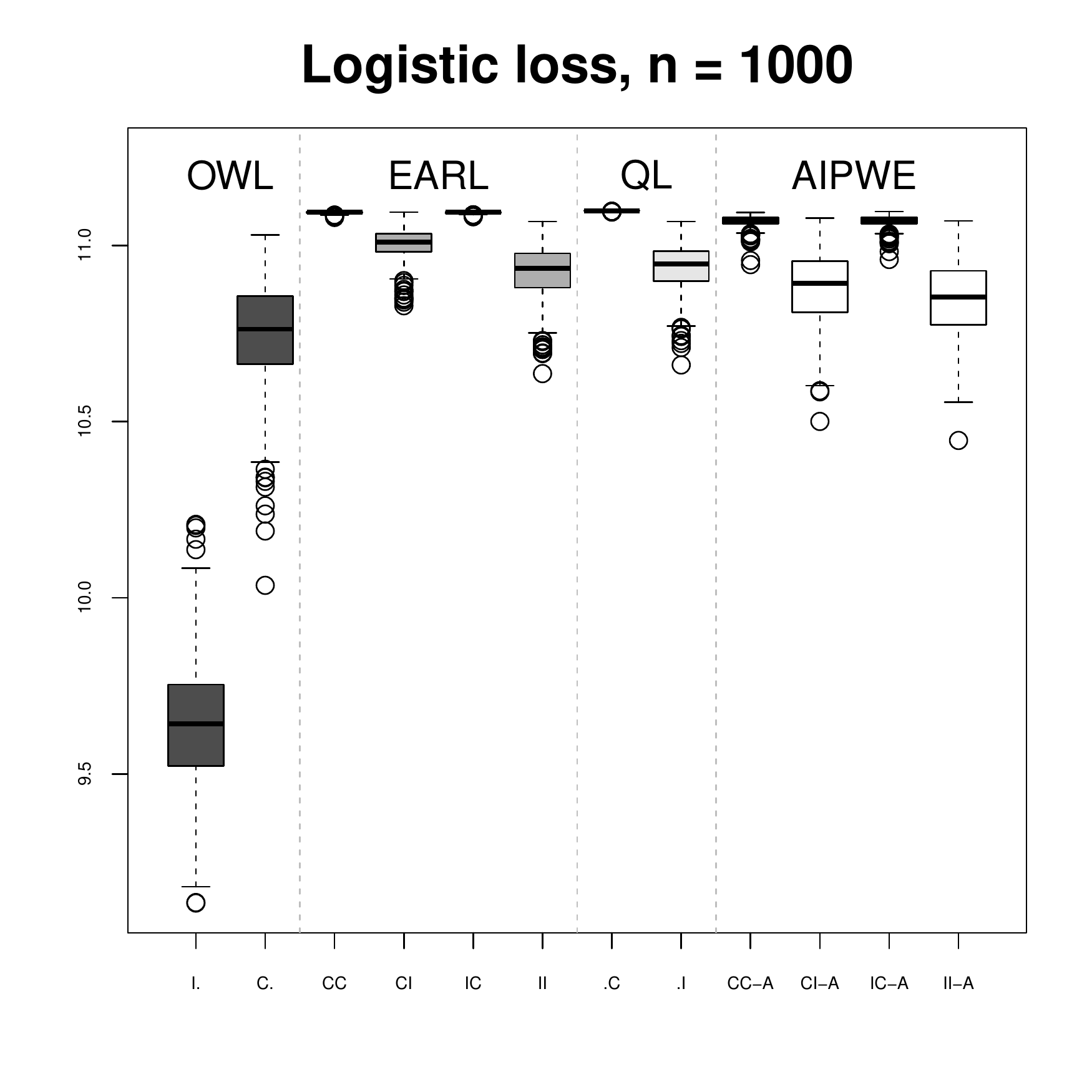}
\includegraphics[width=2.8in,height=2.8in]{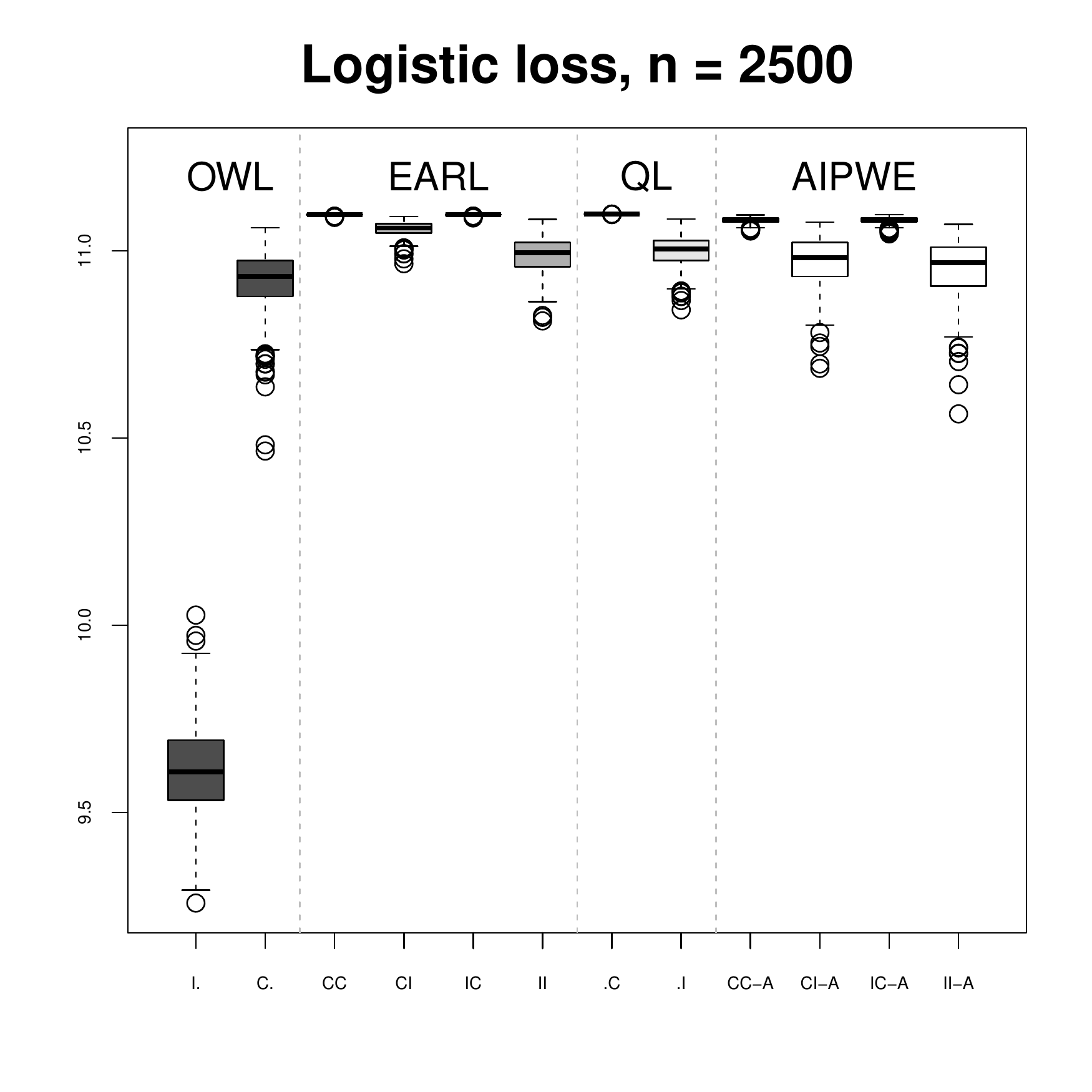}
\includegraphics[width=2.8in,height=2.8in]{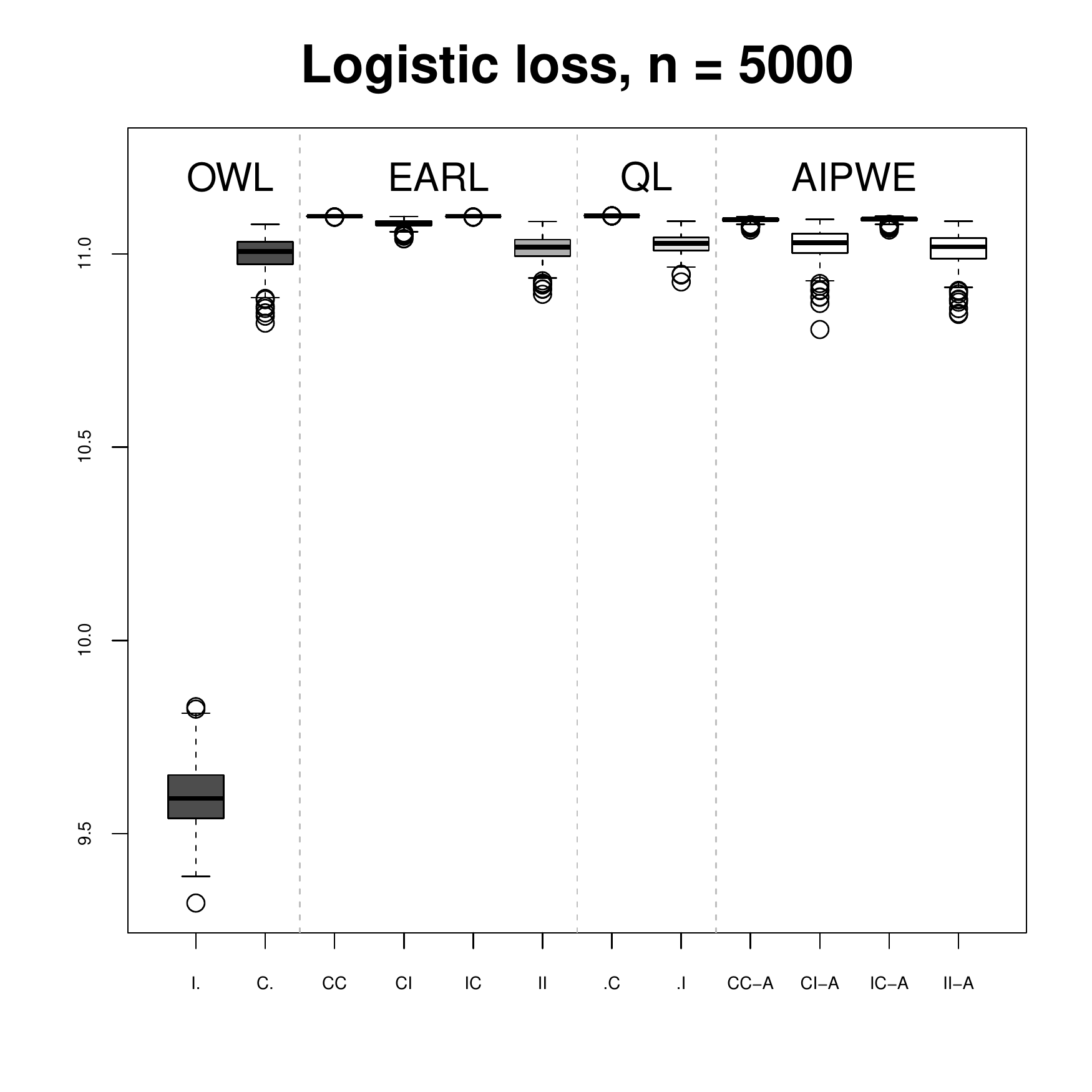}
\includegraphics[width=2.8in,height=2.8in]{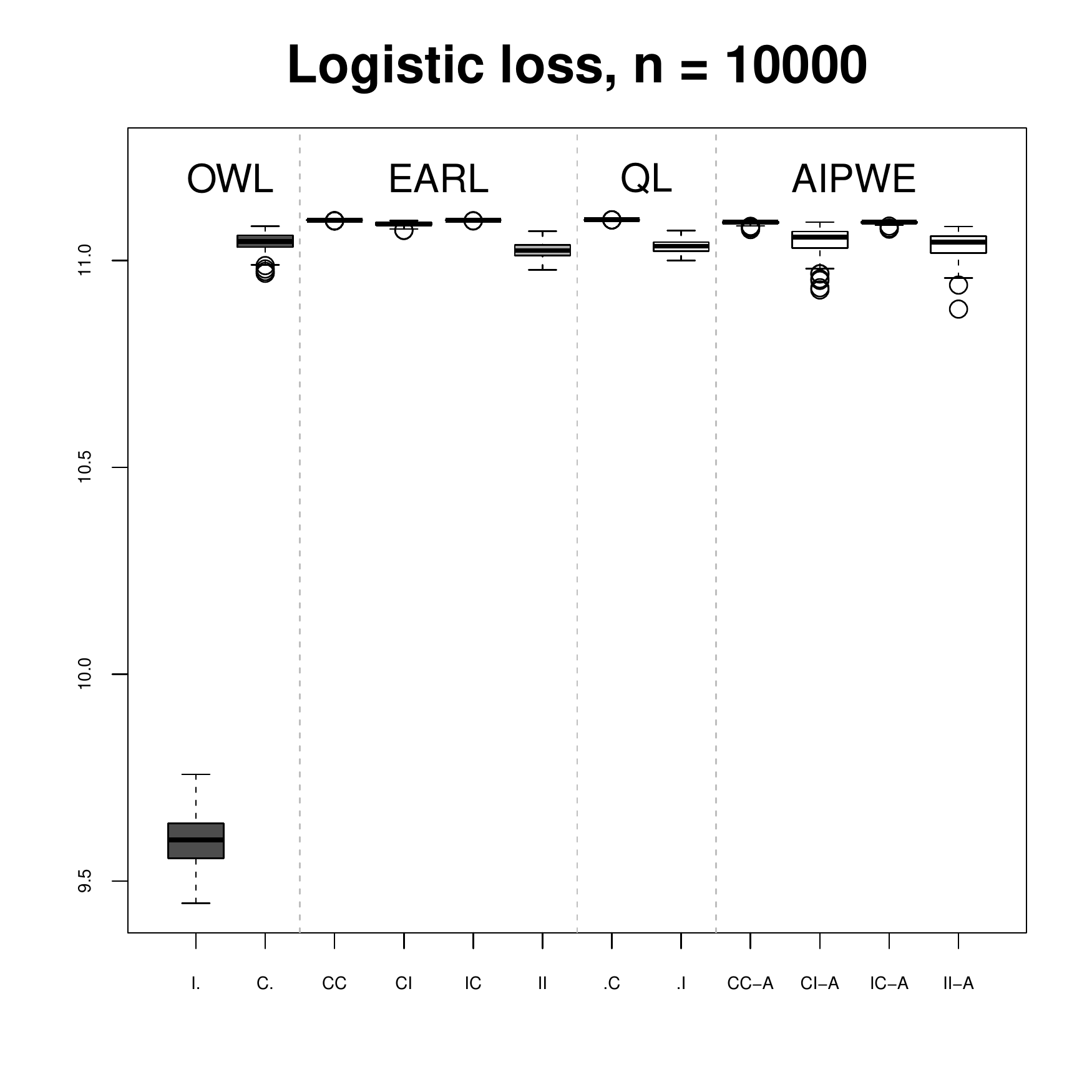}
\end{center}
\end{figure}

 \begin{figure}[h!]
\caption{Boxplots for Scenario 3 results under QL, AIPWE, and OWL and EARL using logistic loss.}
\label{figIPWvsAIPW2}
\begin{center}
 
  \includegraphics[width=2.8in,height=2.8in]{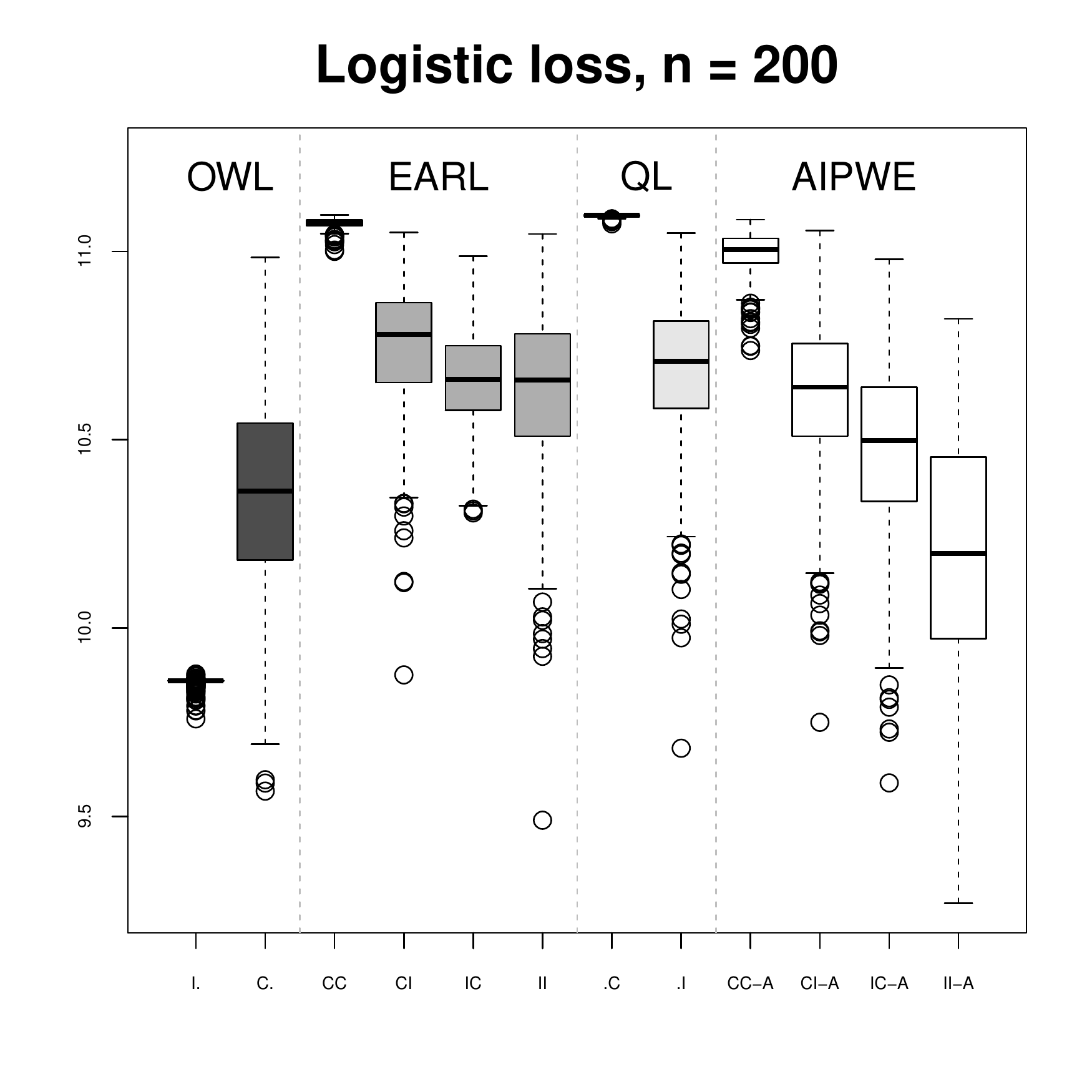}
\includegraphics[width=2.8in,height=2.8in]{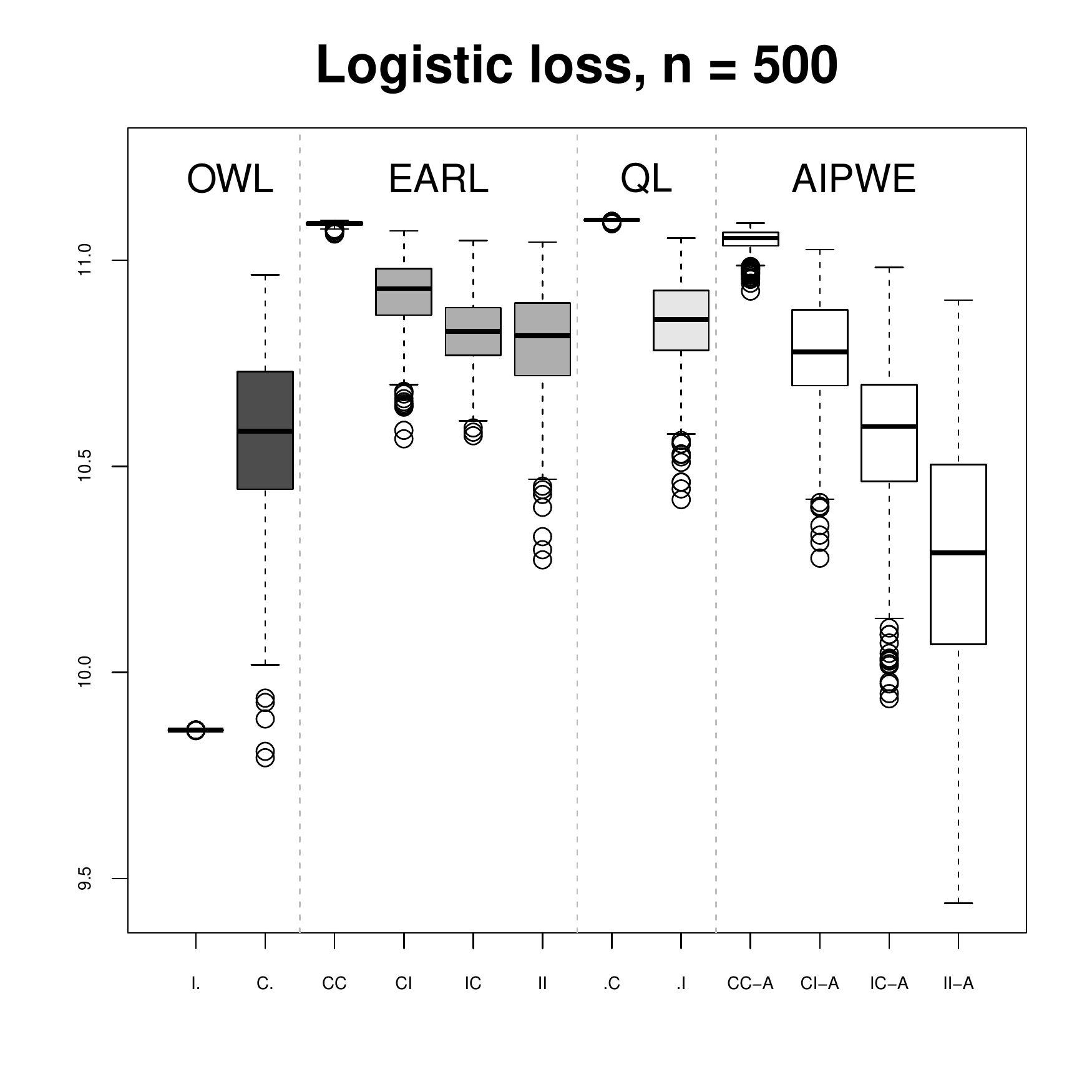}
\includegraphics[width=2.8in,height=2.8in]{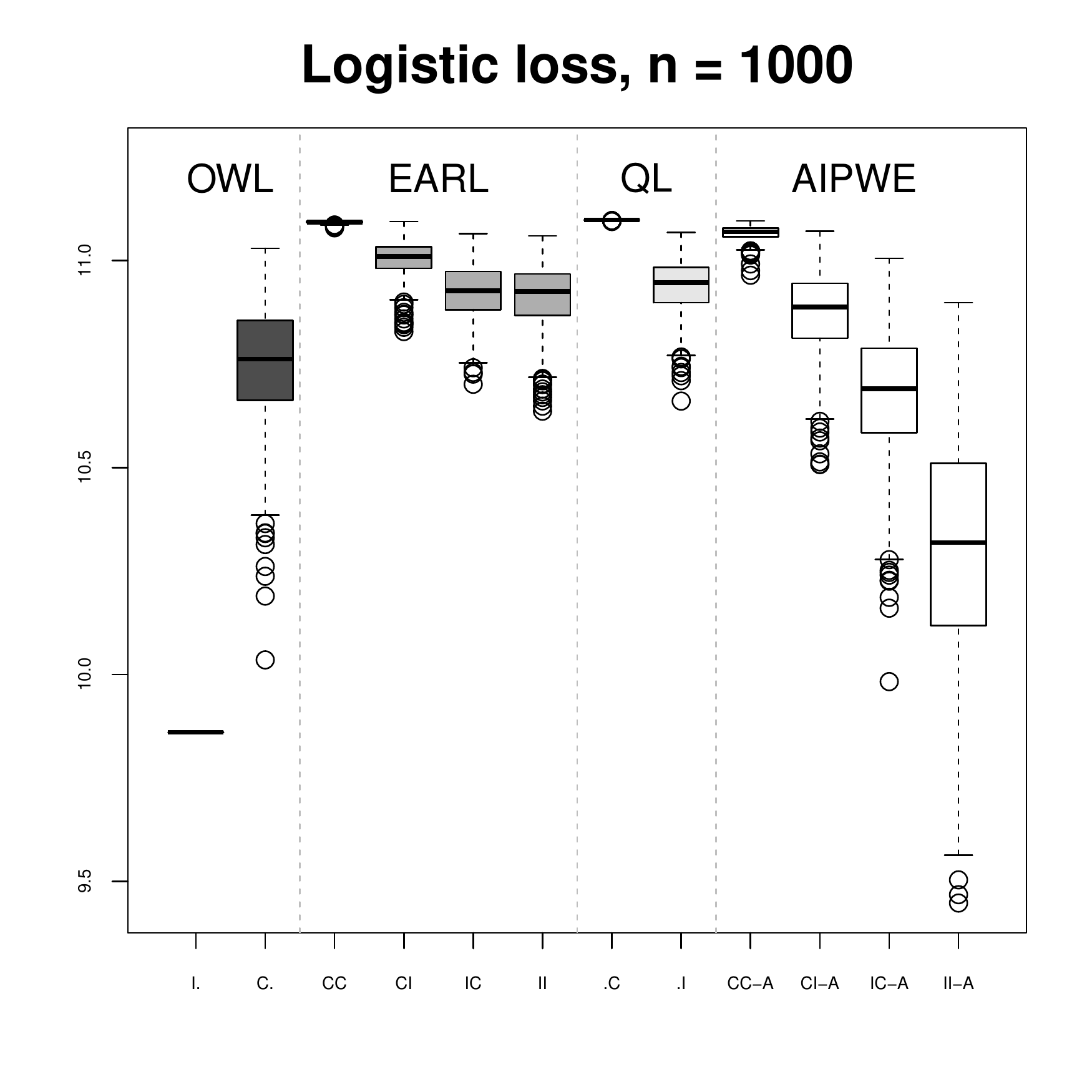}
\includegraphics[width=2.8in,height=2.8in]{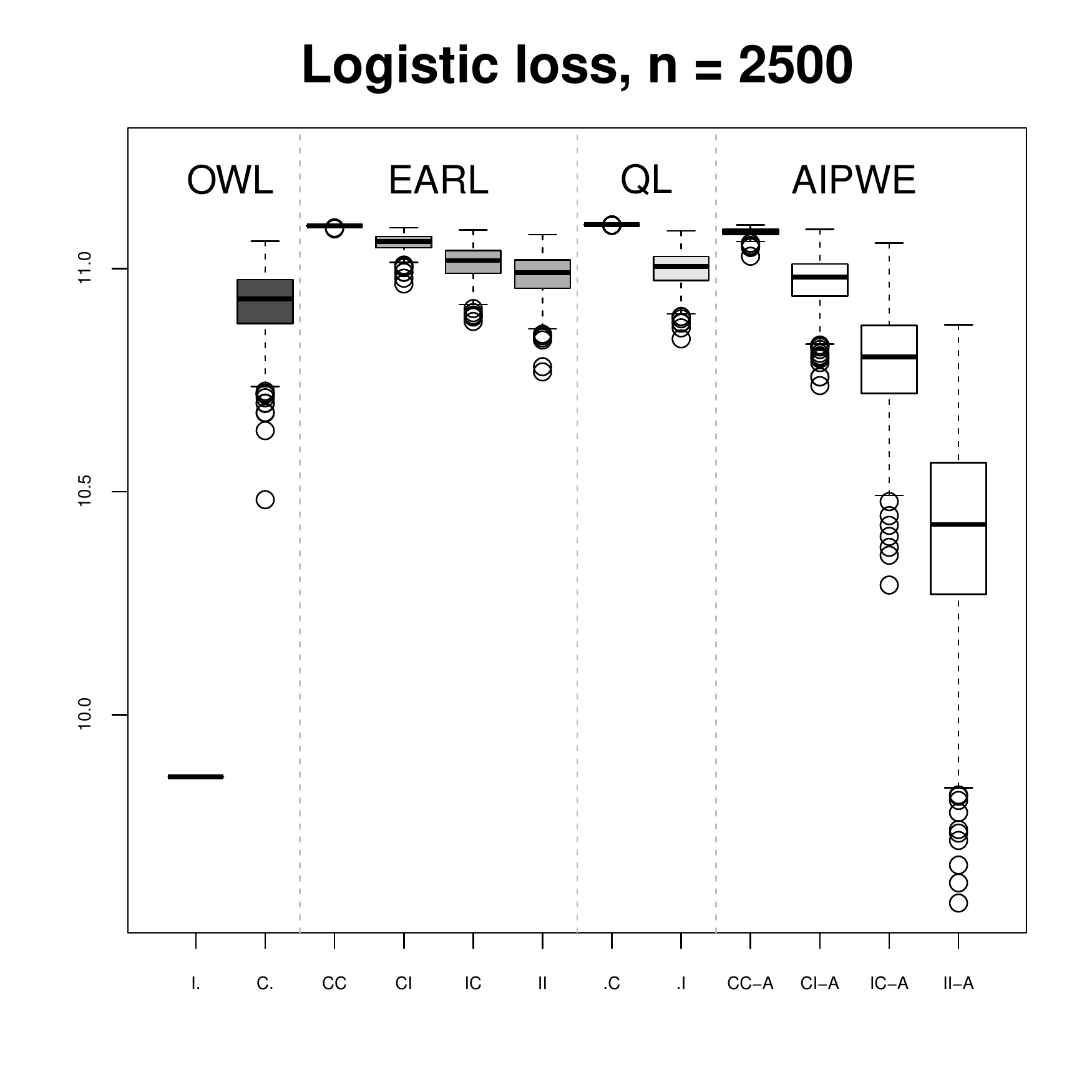}
\includegraphics[width=2.8in,height=2.8in]{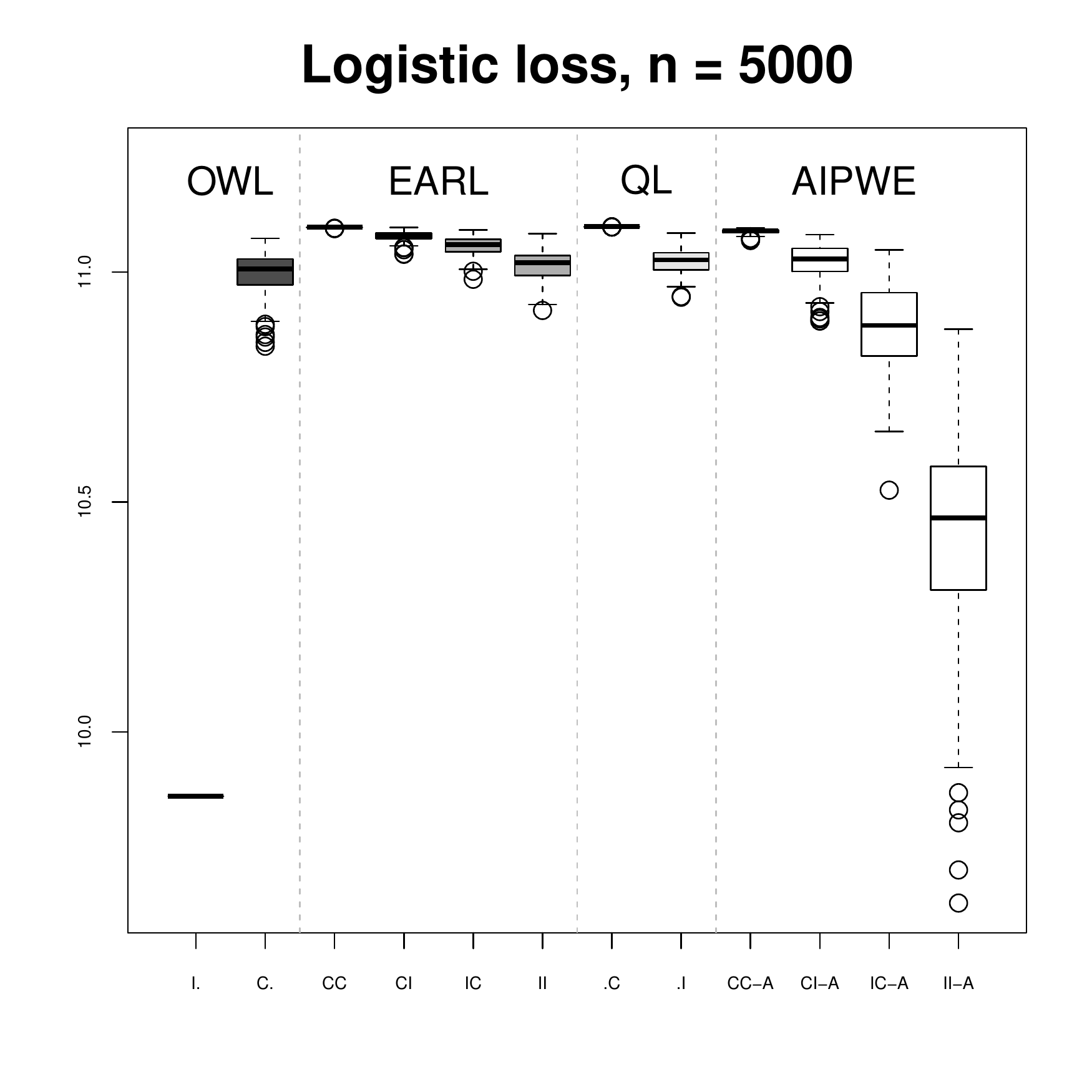}
\includegraphics[width=2.8in,height=2.8in]{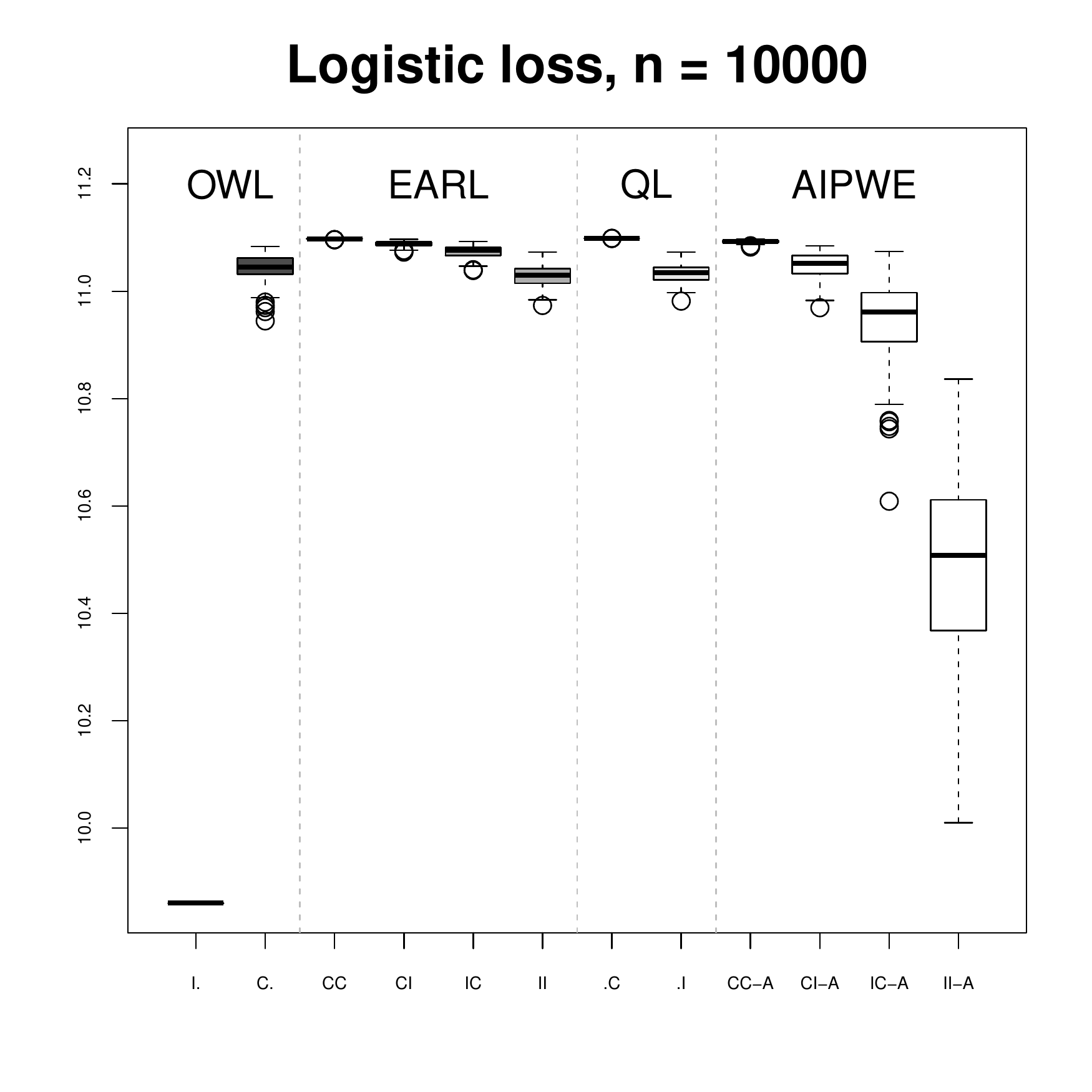}
\end{center}
\end{figure}

It appears that misspecification of the model for the $Q$-function
has a bigger impact than misspecification of the propensity score
model on the AIPWE and EARL methods.
%as well as the AIPWE method, in terms of both biases and
%variances. 
%This is mainly due to the small sample size in estimating
%the propensity score. Such a phenomenon is alleviated as sample size
%increases. 
The relatively poor performance when the propensity is
correctly specified but the regression model is not might be
attributed in part to inverse weighting by the propensity score,
which is problematic when some estimated propensity scores are close
to zero, yielding large weights and subsequently induces bias
\citep{kang2007demystifying}. 
%For example, in Scenario 2, propensity
%scores are bounded away for under the generative mechanism. The
%results are better compared to Scenario 1 under the same regression
%model specification.  
This is illustrated by contrasting scenarios 1 and 2. Propensity scores in Scenario 2 are bounded away from zero, which yield a better result compared to Scenario 1. 
Furthermore, the large variability when the
regression model is misspecified may be partly a consequence of the
method used to estimate the coefficients in the regression model
\citep[see][]{cao2009improving}. 

Finally, we consider an example to illustrate  the impact of a severely misspecified propensity score model. In Scenario 3, the data was generated as in 
Scenario 2 except that the propensity score was set to 0.025 for all subjects. The `CI' setup outperformed the `IC' setup, especially when the sample size was small. Furthermore, the performance of the AIPWE method was largely affected by this poorly imposed propensity model. The results of `CI' and `II'  setups were unsatisfactory even when the sample size was increased to 10000. This example indicates that the performances in the `CI' and `IC' setups depend on the degree of misspecification in the outcome regression model and propensity score model.
%\begin{enumerate} 
%\item[Model Set 1.]  An incorrectly specified logistic regression  model for $\pi(A; \bX)$ with predictors $\bX$ in Scenario 1, and  without any predictors in Scenario 2; an incorrectly specified model for $Q(\bX,A)$ with predictors $\bX, A, \bX A$.
% \item[Model Set 2.] An incorrectly specified logistic regression model for $\pi(A; \bX)$ with predictors $X_1X_2$, and with $X_1^2$ in Scenario 2;  an incorrectly specified model for $Q(\bX,A)$ with predictors $\bX^2, A, \bX A$.
%\end{enumerate}

% \begin{figure}[h!]
%\caption{Boxplots for Scenario 1 results under different model specifications in EARL and AIPWE using logistic loss.}
%\label{incorrect_models1}
%\begin{center}
%  \includegraphics[width=5.2in,height=4.2in]{AIPW-value-logit-incorrect-n-800_SN1.pdf}
% \end{center}
%\end{figure} 

 %\begin{figure}[h!]
%\caption{Boxplots for Scenario 2 results under different model specifications in EARL and AIPWE using logistic loss.}
%\label{incorrect_models2}
%\begin{center}
%  \includegraphics[width=5.2in,height=4.2in]{AIPW-value-logit-incorrect-n-800_SN2.pdf}
% \end{center}
%\end{figure} 
  
%The results from Figures \ref{incorrect_models1} and \ref{incorrect_models2} indicate that  the results of CI and II are improved when  $Q(\bX,A)$ is fitted with $\bX^2, A, \bX A$, although misspecified, compared with the case when  $Q(\bX,A)$ is fitted with $\bX, A, \bX A$.

We also conducted a set of simulation experiments to investigate the role of parametric and nonparametric models for the propensity score and outcome regression. In addition, we compared the performance across different surrogate loss functions, including logistic loss, exponential loss, squared hinge loss, and hinge loss. These additional simulation results can be found in Web Appendix F. In summary, we found that in the examples considered, using nonparametric working models for propensity scores could improve results over parametric models. Hinge loss has a more robust performance when the regression model is incorrect compared to other smooth losses.

%\begin{table}[h]
%\caption{Mean proportion of obtaining optimal treatments}
%\begin{center}
%  \begin{tabular}{ccccccccccc}  
%                &    No Aug       &   P1 & P2 & P3 & P4 & P5 & P6 & P7 & P8 & P9\\
%   n=100   &  0.705 &   0.910 & 0.868 & 0.912 & 0.875 & 0.827 & 0.827 & 0.865 & 0.905 & 0.823  \\       
%  \end{tabular}
%  \label{tb_sim1}
%  \end{center}
%\end{table}

\section{Application: Ocean State Crohn's and Colitis Area Registry  (OSCCAR)}\label{scn:data}
OSCCAR is a community-based incident cohort of subjects with
inflammatory bowel disease (IBD) residing in the state of Rhode Island
that was established in 2008 \citep{sands2009osccar}. Subjects
enrolled in OSCCAR have ulcerative colitis (UC), Crohn's disease (CD),
or indeterminate colitis (IC). Corticosteroids are commonly used to
treat active symptoms. Although, corticosteroids often promptly 
achieve remission, long-term use is complicated by many potential side
effects. One treatment strategy for IBD patients is a ``step-up''
approach in which patients are prescribed medications with increasing
potential toxicity based on the severity of their disease.
Alternatively, a ``top-down'' approach uses aggressive therapy early
in the disease course to prevent long-term complications.  
Both approaches have been shown to be clinically effective, however,
%The
%effectiveness of both approaches has been evaluated clinically. 
there is
treatment response heterogeneity and it is not clear which treatment
is right for each individual patient.  Clinical theory dictates
that those likely to experience a more aggressive disease progression
would benefit more from ``top-down'' than ``step-up''; whereas those
likely to experience a less aggressive progression might benefit more
from ``step-up.''

The primary outcome is the disease activity score measured at the end
of the second year, as measured by the Harvey--Bradshaw Index for
subjects with CD and the Simple Clinical Colitis Index for subjects
with UC.  In both measures, higher scores reflect more disease
activity.  A high-quality treatment rule would reduce disease activity
by assigning patients to top-down if it is necessary and step-up
otherwise.  Among the 274 patients included in the observed data,
32 patients were assigned to the top-down strategy $(A = 1)$ and
242 were assigned to step-up $(A=-1)$.  To remain consistent with our
paradigm of maximizing mean response we used the negative disease
activity score as the response, $Y$. 11 patient covariates were used, which included age, gender, ethnicity, marital
status, race, body mass index, disease type, antibiotics drug usage,
antidiarrheal drug usage, indicator for extra-intestinal
manifestation and baseline disease activity
scores.  We used a linear regression
model to estimate the $Q$-function, and a regularized logistic regression model to
estimate the propensity score to avoid overfitting.  %(see the Web Supplement for a complete description of these models)
In addition to the EARL estimators we applied QL and OWL to estimate
an optimal treatment rule. Because this is an observational study with
unknown propensity scores, we evaluated the estimated treatment rules
$\widehat d$ using inverse probability weighting
$ \widehat{V}^{\mathrm{IPWE}}(\widehat d) =\pn \left[YI\left\lbrace A=\widehat
d(\bX)\right\rbrace/
\widehat \pi(A;\bX)\right ]/\pn\left[I\left\lbrace 
A=\widehat d(\bX)\right\rbrace/\widehat\pi(A;\bX)\right], $
where $\widehat \pi$ is the estimated propensity score. Higher values
of $\widehat{V}^{\mathrm{IPWE}}(\widehat d)$, i.e., lower disease
activity scores, indicate a better overall benefit.

%Logistic loss was shown to have a favorable performance among EARL estimators in our simulation study so we apply it here.  
The coefficients of the estimated optimal treatment rules constructed
from EARL with logistic loss are presented in Table \ref{Datacoef}.  A permutation test based on 2000 permutation times was conducted to obtain the p-value for each covariate, which showed that body mass index was significant at 0.05 level and gender was significant at 0.1 level. In general, patients with a more severe disease status at
baseline are likely to benefit from a top-down therapy. This is
consistent with clinical theory as these symptoms are associated with
higher disease severity. 

Table \ref{matchRate} describes the agreement
between the estimated optimal decision rules constructed using different methods, which shows that the rules  estimated using EARL with different loss functions  give quite similar treatment recommendations. In this table, we also present the agreement between the estimated decision rules and the observed treatments. Compared to the observed treatment allocations, the estimated rules encourage more patients to receive top-down therapy, where 161 patients are recommended to top-down treatment by EARL methods with all loss functions, 225 patients are recommended by OWL method using logistic loss and 145 patients are recommended by QL method respectively.  The estimated disease activity
score is   1.75 using logistic loss, compared with 1.80 for the
QL estimator, and 1.75 for OWL using logistic loss. Although the achieved benefit of the ITR yielded  by OWL and EARL were similar, EARL recommended less patients to the more intensive top-down therapy, which could benefit patients by reducing the side effects. The achieved benefits of the derived ITRs were greater than the benefit that was achieved in the observed dataset, where the average disease activity score was 2.24. Since  top-down therapy is relative new in the practice, to be conservative, physicians tend not to provide such therapy to patients.    Our analysis encourages the usage of top-down therapy for a greater benefit, which can be tailored according to individual characteristics. By looking into the relationship between the observed treatment and covariates, we found that in current practice, physicians were more likely to follow top-down therapy while giving out antibiotics and antidiarrheals drugs in patients with Crohn's disease. The  ITRs resulted from EARL, on the other hand, were more likely to recommend  top-down therapy  for ulcerative colitis/indeterminate colitis patients while they are not taking antibiotics and antidiarrheals drugs. 
%Hence, intensive therapies will control the disease
%progression and relieve the symptoms. Patients who have mild disease
%can use less intensive agents to control the disease activity.

\begin{table}
\caption{Coefficients for the estimated optimal decision rules by EARL with logistic loss (*: significant at 0.05 level).}
\label{Datacoef}
\begin{center}
\begin{tabular}{rcl}
\hline\hline
     & Coefficient  & p-value\\
\hline
Intercept & 2.466  &  \quad- \\
Age &  -0.001 & 0.905\\
Gender (Male = 1) & 0.756 & 0.015$^*$\\ 
Ethnicity (Hispanic = 1) & -1.045 & 0.144\\
Marital status (Single = 1)& -0.320 & 0.318\\
Race (White = 1) & -0.233 & 0.478\\
Body mass index &  -0.063 & 0.037$^{*}$\\
Disease type (UC or IC = 1)& 0.309 & 0.234\\
Antibiotics drug usage (Yes = 1)& -0.156 & 0.563\\
Antidiarrheals drug usage (Yes = 1)& -0.580 & 0.167\\
Extra-intestinal manifestation  (Yes = 1)& 0.273 & 0.286\\
Baseline disease activity scores  & 0.050 & 0.427\\  
\hline
\hline
\end{tabular}
\end{center}
\end{table}

\begin{table}
\begin{center}
\caption{Agreements between the estimated optimal decision rule yield by different methods and the observed treatment. OWL-logit: OWL using logistic loss; EARL: EARL using logistic loss; EARL: EARL using exponential oss; EARL-hinge: EARL using hinge loss; EARL-sqhinge: EARL using squared hinge loss; QL: Q-learning.}
\label{matchRate}
\vspace{1em}
{\small \begin{tabular}{rccccccc}
\hline
   & OWL-Logit & EARL-logit & EARL-exp & EARL-hinge & EARL-sqhinge & QL & \\ 
   \hline
 OWL-Logit   &  1  &  0.642  & 0.821  & 0.639 & 0.639 &    0.577  \\
EARL-logit    &    & 1 & 0.588 &   0.996 & 0.996   &  0.920  \\
 EARL-exp    &     &  & 1 & 0.591 & 0.591 &  0.529 \\
 EARL-hinge  &    &  &    &   1     &  1 & 0.916 \\
  EARL-sqhinge &   &   &    &      & 1 & 0.916 \\
  QL &   &   &    &      &   & 1 \\
  \hline
  Observed & 0.193 & 0.449 & 0.117 & 0.453 & 0.453 & 0.507\\
\hline  
\end{tabular}}
\end{center}
\end{table}

 We also applied our method to the study of National Supported Work Demonstration, which also showed a superior performance of the proposed method. Results are shown in Web Appendix G. 

\section{Discussion}
We proposed a class of estimators for the optimal treatment
rule that we termed EARL.  This class of methods is
formed by applying a convex relaxation to the AIPWE
of the marginal mean outcome.  To reduce the 
risk of misspecification, it is possible to use flexible,
e.g., nonparametric, models for the propensity score and
 the $Q$-function.  However, we showed theoretically
and empirically that such flexibility comes at the cost of 
additional variability and potentially poor small sample
performance.  

We demonstrated that extreme propensity scores may lead to a large
variance in the augmented inverse probability weighted estimator. To
alleviate this issue, we may consider an estimator which achieves the
smallest variance among its class of doubly robust estimators when the
propensity score model is correctly specified. Such an estimator can
be derived following the techniques used in \cite{cao2009improving}.

There are several important ways  this work might be extended.  The
first is to handle time-to-event outcomes wherein the observed data
are subject to right-censoring.  In this setting, efficient methods for
augmentation to adjust for censoring might be folded into the EARL
framework.   Another extension is to multi-stage treatment rules, also known as, dynamic treatment
regimes \citep{Murphy03optimaldynamic, Robins04optimalstructural,
  moodie:optimaldynamic07}.  A challenging component
of this extension is that the variability of the AIPWE increases
dramatically as the number of treatment stages increases.  We believe
that the convex relaxation may help in this setting not only in terms
of computation but also by reducing variance. 
%We are currently
%exploring this extension.
%\fi

\section{Supplementary Materials}
The Web Appendix referenced in Sections \ref{scntheory}, \ref{simSection} and \ref{scn:data} is available online.

 \bibliographystyle{jasa}
\bibliography{DRobservational}

\begin{thebibliography}{51}
\providecommand{\natexlab}[1]{#1}
\providecommand{\url}[1]{\texttt{#1}}
\expandafter\ifx\csname urlstyle\endcsname\relax
  \providecommand{\doi}[1]{doi: #1}\else
  \providecommand{\doi}{doi: \begingroup \urlstyle{rm}\Url}\fi

\bibitem[Athey and Wager(2017)]{athey2017efficient}
Susan Athey and Stefan Wager.
\newblock Efficient policy learning.
\newblock \emph{arXiv preprint arXiv:1702.02896}, 2017.

\bibitem[Bang and Robins(2005)]{bang2005doubly}
Heejung Bang and James~M Robins.
\newblock Doubly robust estimation in missing data and causal inference models.
\newblock \emph{Biometrics}, 61\penalty0 (4):\penalty0 962--973, 2005.

\bibitem[Bartlett et~al.(2006)Bartlett, Jordan, and
  McAuliffe]{bartlett:convexity06}
Peter~L Bartlett, Michael~I Jordan, and Jon~D McAuliffe.
\newblock Convexity, classification, and risk bounds.
\newblock \emph{J. of {A}merican {S}tatistical {A}ssociation}, 101\penalty0
  (473):\penalty0 138--156, 2006.

\bibitem[Benkeser et~al.(2017)Benkeser, Carone, Laan, and
  Gilbert]{benkeser2017doubly}
David Benkeser, Marco Carone, MJ~Van~Der Laan, and PB~Gilbert.
\newblock Doubly robust nonparametric inference on the average treatment
  effect.
\newblock \emph{Biometrika}, 104\penalty0 (4):\penalty0 863--880, 2017.

\bibitem[Bickel(1982)]{bickel1982adaptive}
Peter~J Bickel.
\newblock On adaptive estimation.
\newblock \emph{The Annals of Statistics}, pages 647--671, 1982.

\bibitem[Cao et~al.(2009)Cao, Tsiatis, and Davidian]{cao2009improving}
Weihua Cao, Anastasios~A Tsiatis, and Marie Davidian.
\newblock Improving efficiency and robustness of the doubly robust estimator
  for a population mean with incomplete data.
\newblock \emph{Biometrika}, 96\penalty0 (3):\penalty0 723--734, 2009.

\bibitem[Chakraborty and Moodie(2013)]{bibhasBook}
Bibhas Chakraborty and Erica~EM Moodie.
\newblock \emph{Statistical Methods for Dynamic Treatment Regimes}.
\newblock Springer, 2013.

\bibitem[Chernozhukov et~al.(2016)Chernozhukov, Chetverikov, Demirer, Duflo,
  Hansen, and Newey]{chernozhukov2016double}
Victor Chernozhukov, Denis Chetverikov, Mert Demirer, Esther Duflo, Christian
  Hansen, and Whitney~K Newey.
\newblock Double machine learning for treatment and causal parameters.
\newblock Technical report, cemmap working paper, Centre for Microdata Methods
  and Practice, 2016.

\bibitem[Davidian et~al.(2014)Davidian, Tsiatis, and Laber]{marieChapter}
M.~Davidian, A.A. Tsiatis, and E.B. Laber.
\newblock Value search estimators.
\newblock In \emph{Dynamic Treatment Regimes}, pages 1--40. Springer, 2014.

\bibitem[Fan et~al.(2016)Fan, Imai, Liu, Ning, and Yang]{fan2016improving}
Jianqing Fan, Kosuke Imai, Han Liu, Yang Ning, and Xiaolin Yang.
\newblock Improving covariate balancing propensity score: A doubly robust and
  efficient approach.
\newblock Technical report, 2016.

\bibitem[Freund and Schapire(1999)]{freund1999large}
Yoav Freund and Robert~E Schapire.
\newblock Large margin classification using the perceptron algorithm.
\newblock \emph{Machine learning}, 37\penalty0 (3):\penalty0 277--296, 1999.

\bibitem[Hastie et~al.(2009)Hastie, Tibshirani, and Friedman]{hastie:esl09}
T.~Hastie, R.~Tibshirani, and J.~H. Friedman.
\newblock \emph{The Elements of Statistical Learning}.
\newblock Springer-Verlag New York, Inc., New York, second edition, 2009.

\bibitem[Henderson et~al.(2009)Henderson, Ansell, and
  Alshibani]{henderson2009regret}
R.~Henderson, P.~Ansell, and D.~Alshibani.
\newblock {Regret-Regression for Optimal Dynamic Treatment Regimes}.
\newblock \emph{Biometrics}, 66\penalty0 (4), 2009.

\bibitem[Kang and Schafer(2007)]{kang2007demystifying}
Joseph~DY Kang and Joseph~L Schafer.
\newblock Demystifying double robustness: A comparison of alternative
  strategies for estimating a population mean from incomplete data.
\newblock \emph{Statistical science}, pages 523--539, 2007.

\bibitem[Kitagawa and Tetenov(2017)]{kitagawa2017should}
Toru Kitagawa and Aleksey Tetenov.
\newblock Who should be treated? empirical welfare maximization methods for
  treatment choice.
\newblock 2017.

\bibitem[Kosorok(2008)]{kosorok:ep08}
M.~R. Kosorok.
\newblock \emph{Introduction to empirical processes and semiparametric
  inference}.
\newblock Springer-Verlag, New York, 2008.

\bibitem[Laber and Murphy(2011)]{laber2011adaptive}
Eric~B Laber and Susan~A Murphy.
\newblock Adaptive confidence intervals for the test error in classification.
\newblock \emph{Journal of the American Statistical Association}, 106\penalty0
  (495):\penalty0 904--913, 2011.

\bibitem[Laber et~al.(2014)Laber, Lizotte, Qian, Pelham, Murphy,
  et~al.]{laber2014dynamic}
Eric~B Laber, Daniel~J Lizotte, Min Qian, William~E Pelham, Susan~A Murphy,
  et~al.
\newblock Dynamic treatment regimes: Technical challenges and applications.
\newblock \emph{Electronic Journal of Statistics}, 8:\penalty0 1225--1272,
  2014.

\bibitem[Linn et~al.(2016)Linn, Laber, and Stefanski]{linn2016interactive}
Kristin~A Linn, Eric~B Laber, and Leonard~A Stefanski.
\newblock Interactive q-learning for quantiles.
\newblock \emph{Journal of the American Statistical Association}, \penalty0
  (just-accepted):\penalty0 1--37, 2016.

\bibitem[Liu et~al.(2016)Liu, Wang, Kosorok, Zhao, and Zeng]{liu2016robust}
Ying Liu, Yuanjia Wang, Michael~R Kosorok, Yingqi Zhao, and Donglin Zeng.
\newblock Robust hybrid learning for estimating personalized dynamic treatment
  regimens.
\newblock \emph{arXiv preprint arXiv:1611.02314}, 2016.

\bibitem[Moodie et~al.(2007)Moodie, Richardson, and
  Stephens]{moodie:optimaldynamic07}
Erica E.~M. Moodie, Thomas~S. Richardson, and David~A. Stephens.
\newblock Demystifying optimal dynamic treatment regimes.
\newblock \emph{Biometrics}, 63\penalty0 (2):\penalty0 447--455, 2007.

\bibitem[Moodie et~al.(2012)Moodie, Chakraborty, and Kramer]{moodie2012q}
Erica~EM Moodie, Bibhas Chakraborty, and Michael~S Kramer.
\newblock Q-learning for estimating optimal dynamic treatment rules from
  observational data.
\newblock \emph{Canadian Journal of Statistics}, 40\penalty0 (4):\penalty0
  629--645, 2012.

\bibitem[Moodie et~al.(2013)Moodie, Dean, and Sun]{moodie2013q}
Erica~EM Moodie, Nema Dean, and Yue~Ru Sun.
\newblock Q-learning: Flexible learning about useful utilities.
\newblock \emph{Statistics in Biosciences}, pages 1--21, 2013.

\bibitem[Murphy(2003)]{Murphy03optimaldynamic}
S.~A. Murphy.
\newblock Optimal dynamic treatment regimes.
\newblock \emph{Journal of the Royal Statistical Society, Series B},
  65:\penalty0 331--366, 2003.

\bibitem[Newey(1997)]{newey1997convergence}
Whitney~K Newey.
\newblock Convergence rates and asymptotic normality for series estimators.
\newblock \emph{Journal of econometrics}, 79\penalty0 (1):\penalty0 147--168,
  1997.

\bibitem[Orellana et~al.(2010)Orellana, Rotnitzky, and Robins]{Orellana10}
L.~Orellana, A.~Rotnitzky, and J.~Robins.
\newblock Dynamic regime marginal structural mean models for estimation of
  optimal dynamic treatment regimes, part i: Main content.
\newblock \emph{Int. Jrn. of Biostatistics}, 6\penalty0 (2):\penalty0 1--19,
  2010.

\bibitem[Qian and Murphy(2011)]{qian:itr11}
Min Qian and S.~A. Murphy.
\newblock Performance guarantees for individualized treatment rules.
\newblock \emph{The Annals of Statistics}, 39:\penalty0 1180--1210, 2011.

\bibitem[Robins(1986)]{Robins:Causal1986}
James Robins.
\newblock A new approach to causal inference in mortality studies with a
  sustained exposure period—application to control of the healthy worker
  survivor effect.
\newblock \emph{Mathematical Modelling}, 7:\penalty0 1393--1512, 1986.

\bibitem[Robins(1997)]{Robins:CausalNotes1997}
James Robins.
\newblock Causal inference from complex longitudinal data.
\newblock \emph{Lect. Notes Statist.}, 120:\penalty0 69--117, 1997.

\bibitem[Robins(1989)]{robins1989analysis}
James~M Robins.
\newblock The analysis of randomized and non-randomized aids treatment trials
  using a new approach to causal inference in longitudinal studies.
\newblock \emph{Health service research methodology: a focus on AIDS},
  113:\penalty0 159, 1989.

\bibitem[Robins(2004)]{Robins04optimalstructural}
James~M. Robins.
\newblock Optimal structural nested models for optimal sequential decisions.
\newblock In \emph{In Proceedings of the Second Seattle Symposium on
  Biostatistics}, pages 189--326. Springer, 2004.

\bibitem[Robins et~al.(1994)Robins, Rotnitzky, and Zhao]{robins1994estimation}
James~M Robins, Andrea Rotnitzky, and Lue~Ping Zhao.
\newblock Estimation of regression coefficients when some regressors are not
  always observed.
\newblock \emph{Journal of the American Statistical Association}, 89\penalty0
  (427):\penalty0 846--866, 1994.

\bibitem[Robins et~al.(2017)Robins, Li, Mukherjee, Tchetgen, van~der Vaart,
  et~al.]{robins2017minimax}
James~M Robins, Lingling Li, Rajarshi Mukherjee, Eric~Tchetgen Tchetgen, Aad
  van~der Vaart, et~al.
\newblock Minimax estimation of a functional on a structured high-dimensional
  model.
\newblock \emph{The Annals of Statistics}, 45\penalty0 (5):\penalty0
  1951--1987, 2017.

\bibitem[Robins et~al.(2008)Robins, Orellana, and Rotnitzky]{robinsetal2008}
J.M. Robins, L.~Orellana, and A.~Rotnitzky.
\newblock Estimation and extrapolation of optimal treatment and testing
  strategies.
\newblock \emph{Statistics in Medicine}, pages 4678--4721, 2008.

\bibitem[Rubin(1974)]{Rubin:Causal1974}
D.~B. Rubin.
\newblock Estimating causal effects of treatments in randomized and
  nonrandomized studies.
\newblock \emph{Journal of Educational Psychology}, 66:\penalty0 688--701,
  1974.

\bibitem[Sands et~al.(2009)Sands, LeLeiko, Shah, Bright, and
  Grabert]{sands2009osccar}
Bruce~E Sands, Neal LeLeiko, Samir~A Shah, Renee Bright, and Stacey Grabert.
\newblock Osccar: ocean state crohn's and colitis area registry.
\newblock \emph{Medicine and Health Rhode Island}, 92\penalty0 (3):\penalty0
  82, 2009.

\bibitem[Schick(1986)]{schick1986asymptotically}
Anton Schick.
\newblock On asymptotically efficient estimation in semiparametric models.
\newblock \emph{The Annals of Statistics}, pages 1139--1151, 1986.

\bibitem[Schulte et~al.(2014)Schulte, Tsiatis, Laber, , and
  Davidian]{Schulte:QandA2012}
Phillip~J. Schulte, Anastasios~A. Tsiatis, Eric~B. Laber, , and Marie Davidian.
\newblock Q- and a-learning methods for estimating optimal dynamic treatment
  regimes.
\newblock \emph{Statistical Science}, 29:\penalty0 640--661, 2014.

\bibitem[Sox and Greenfield(2009)]{sox2009comparative}
Harold~C Sox and Sheldon Greenfield.
\newblock Comparative effectiveness research: a report from the institute of
  medicine.
\newblock \emph{Annals of Internal Medicine}, 151\penalty0 (3):\penalty0
  203--205, 2009.

\bibitem[Splawa-Neyman et~al.(1990)Splawa-Neyman, Dabrowska, and
  Speed]{Neyman:PotentialOutcome1990}
J.~Splawa-Neyman, DM~Dabrowska, and TP~Speed.
\newblock On the application of probability theory to agricultural experiments
  (engl. transl. by d.m. dabrowska and t.p. speed).
\newblock \emph{Statistical Science}, 5:\penalty0 465--472, 1990.

\bibitem[Sutton and Barto(1998)]{Sutton:Reinforcementlearning98}
Richard~S. Sutton and Andrew~G. Barto.
\newblock \emph{Reinforcement Learning I: Introduction}.
\newblock MIT Press, Cambridge,MA, 1998.

\bibitem[Szepesv{\'a}ri(2010)]{szepesvari2010algorithms}
Csaba Szepesv{\'a}ri.
\newblock Algorithms for reinforcement learning.
\newblock \emph{Synthesis Lectures on Artificial Intelligence and Machine
  Learning}, 4\penalty0 (1):\penalty0 1--103, 2010.

\bibitem[Taylor et~al.(2015)Taylor, Cheng, and Foster]{taylor2015reader}
Jeremy~MG Taylor, Wenting Cheng, and Jared~C Foster.
\newblock Reader reaction to ga robust method for estimating optimal treatment
  regimesh by zhang et al.(2012).
\newblock \emph{Biometrics}, 71\penalty0 (1):\penalty0 267--273, 2015.

\bibitem[Zhang et~al.(2012{\natexlab{a}})Zhang, Tsiatis, Davidian, Zhang, and
  Laber]{baqun2012}
Baqun Zhang, Anastasios~A Tsiatis, Marie Davidian, Min Zhang, and Eric Laber.
\newblock Estimating optimal treatment regimes from a classification
  perspective.
\newblock \emph{Stat}, 1\penalty0 (1):\penalty0 103--114, 2012{\natexlab{a}}.

\bibitem[Zhang et~al.(2012{\natexlab{b}})Zhang, Tsiatis, Laber, and
  Davidian]{zhang2012robust}
Baqun Zhang, Anastasios~A Tsiatis, Eric~B Laber, and Marie Davidian.
\newblock A robust method for estimating optimal treatment regimes.
\newblock \emph{Biometrics}, 68\penalty0 (4):\penalty0 1010--1018,
  2012{\natexlab{b}}.

\bibitem[Zhang et~al.(2013)Zhang, Tsiatis, Laber, and
  Davidian]{zhang2013robust}
Baqun Zhang, Anastasios~A Tsiatis, Eric~B Laber, and Marie Davidian.
\newblock Robust estimation of optimal dynamic treatment regimes for sequential
  treatment decisions.
\newblock \emph{Biometrika}, 100:\penalty0 681--695, 2013.

\bibitem[Zhang et~al.(2015)Zhang, Laber, Tsiatis, and Davidian]{zhang2015using}
Yichi Zhang, Eric~B Laber, Anastasios Tsiatis, and Marie Davidian.
\newblock Using decision lists to construct interpretable and parsimonious
  treatment regimes.
\newblock \emph{Biometrics}, 71\penalty0 (4):\penalty0 895--904, 2015.

\bibitem[Zhang et~al.(2017)Zhang, Laber, Davidian, and
  Tsiatis]{zhang2017estimation}
Yichi Zhang, Eric~B Laber, Marie Davidian, and Anastasios~A Tsiatis.
\newblock Estimation of optimal treatment regimes using lists.
\newblock \emph{Journal of the American Statistical Association}, \penalty0
  (just-accepted), 2017.

\bibitem[Zhao et~al.(2012)Zhao, Zeng, Rush, and Kosorok]{Zhao:OWL12}
Y.~Q. Zhao, Donglin Zeng, A.~John Rush, and Michael~R. Kosorok.
\newblock Estimating individualized treatment rules using outcome weighted
  learning.
\newblock \emph{Journal of American Statistical Association}, 107:\penalty0
  1106--1118, 2012.

\bibitem[Zhao et~al.(2009)Zhao, Kosorok, and Zeng]{Zhao:RL2009}
Yufan Zhao, Michael~R. Kosorok, and Donglin Zeng.
\newblock Reinforcement learning design for cancer clinical trials.
\newblock \emph{Statistics in Medicine}, 28:\penalty0 3294--3315, 2009.

\bibitem[Zheng and van~der Laan(2011)]{zheng2011cross}
Wenjing Zheng and Mark~J van~der Laan.
\newblock Cross-validated targeted minimum-loss-based estimation.
\newblock In \emph{Targeted Learning}, pages 459--474. Springer, 2011.

\end{thebibliography}

\label{lastpage}

 \end{document}